\documentclass[USenglish,oneside,twocolumn]{article}

\usepackage{etex}
\usepackage[big]{dgruyter_NEW}
\usepackage{enumerate}
\usepackage{epsfig}
\usepackage{graphicx}
\usepackage{epstopdf}
\usepackage{amsmath}
\usepackage{amssymb}
\usepackage{url}
\usepackage{wrapfig}
\usepackage{color}
\usepackage{paralist}
\usepackage{booktabs} 
\usepackage{multirow}
\usepackage{tabularx}
\usepackage{algorithm}
\usepackage{xcolor}
\usepackage{listings}
\usepackage{soul}
\usepackage{placeins}
\usepackage{pgf-umlsd}
\usepackage{tikz}
\usepackage{subfigure}
\usepackage{algorithm}
\usepackage{fixltx2e}

\cclogo{\includegraphics{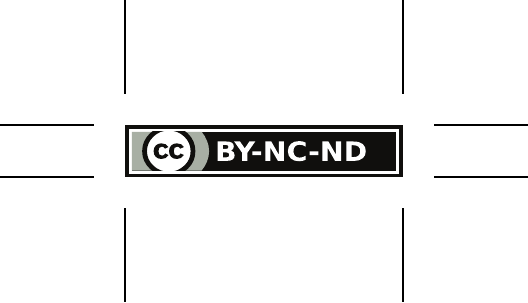}}

\newcommand{\newmaterial}[1]{\textcolor{black}{{#1}}}

\begin{document}

\author*[1]{Ruben Recabarren}
\author[2]{Bogdan Carbunar}
\affil[1]{Florida Int'l University, Miami, FL 33199, E-mail: recabarren@gmail.com}
\affil[2]{Florida Int'l University, Miami, FL 33199, E-mail: carbunar@gmail.com}

\title{\huge{Hardening Stratum, the Bitcoin Pool Mining Protocol}}

\runningtitle{Hardening Stratum, the Bitcoin Pool Mining Protocol}

\begin{abstract}
{Stratum, the de-facto mining communication protocol used by blockchain based
\newmaterial{cryptocurrency} systems, enables miners to reliably and
efficiently fetch jobs from mining pool servers. In this paper we exploit
Stratum's lack of encryption to develop passive and active attacks on Bitcoin's
mining protocol, with important implications on the privacy, security and even
safety of mining equipment owners. We introduce StraTap and ISP Log attacks,
that infer miner earnings if given access to miner communications, or even
their logs. We develop BiteCoin, an active attack that hijacks shares submitted
by miners, and their associated payouts. We build BiteCoin on WireGhost, a tool
we developed to hijack and surreptitiously maintain Stratum connections.\\
Our attacks reveal that securing Stratum through pervasive encryption is not
only undesirable (due to large overheads), but also ineffective: an adversary
can predict miner earnings even when given access to only packet timestamps.
Instead, we devise Bedrock, a minimalistic Stratum extension that protects the
privacy and security of mining participants. We introduce and leverage the {\it
mining cookie} concept, a secret that each miner shares with the pool and
includes in its puzzle computations, and that prevents attackers from
reconstructing or hijacking the puzzles.\\
We have implemented our attacks and collected 138MB of Stratum protocol traffic
from mining equipment in the US and Venezuela. We show that Bedrock is
resilient to active attacks even when an adversary breaks the crypto constructs
it uses. Bedrock imposes a daily overhead of 12.03s on a single pool server
that handles mining traffic from 16,000 miners.}
\end{abstract}


\journalname{Proceedings on Privacy Enhancing Technologies}
\DOI{Editor to enter DOI}
\startpage{1}
\received{..}
\revised{..}
\accepted{..}

\journalyear{..}
\journalvolume{..}
\journalissue{..}

\maketitle

\section{Introduction}
\label{sec:introduction}

\newmaterial{
The privacy and security of Bitcoin have been extensively
studied~\cite{BKP14,KKM14,MPJLMVS13,AKRSC13,MGGR13,BSCG0MTV14,BNMCKF14} and
documented~\cite{BMCNKF15}. While the focus of previous work has been on the
architectural vulnerabilities of the cryptocurrency, no work has been done
to analyze implementation vulnerabilities of the Stratum mining protocol, the
main Bitcoin mining option.}


However, mining activities have important privacy
implications~\cite{MiningSecret}.
%
%
\newmaterial{
Learning the payouts of miners can make them targets of hacking and coin
theft~\cite{BitcoinTheft,ListBitcoinTheft},
kidnapping~\cite{williams2011kidnappings}, and, in countries where Bitcoin is
illegal~\cite{WikiIllegal,BitcoinIllegal}, expose them to arrest and equipment
confiscation~\cite{BitcoinArrests,BitcoinCrackdown}. For instance, in
countries like Venezuela, intelligence police hunt Bitcoin miners to extort,
steal mining equipment or prosecute~\cite{VenezuelaBitcoin}.}

In this paper we study the vulnerabilities of the \textit{Stratum}
protocol~\cite{stratum_proto}, the de facto mining standard for pooled Bitcoin
mining~\cite{stratum_GBT} as well as alternative coins mining, e.g.,
Litecoin~\cite{litecoin_stratum}, Ethereum~\cite{ethereum_stratum} and
Monero~\cite{monero_stratum}; currently the altcoins with the most market
capitalization \cite{crypto_market_cap}. Stratum replaced the original
\textit{getwork} protocol of Bitcoin mining~\cite{stratumDescription}, to
enable miners to fetch jobs from mining pool servers more reliably and
efficiently. In Stratum, the miners solve assigned {\it jobs} and send their
results back in the form of \textit{shares}.  The mining pool server then
compensates the miner according to the difficulty of the assigned jobs and the
number of shares accepted.

\begin{figure}[t!]
\centering
\includegraphics[width=0.49\textwidth]{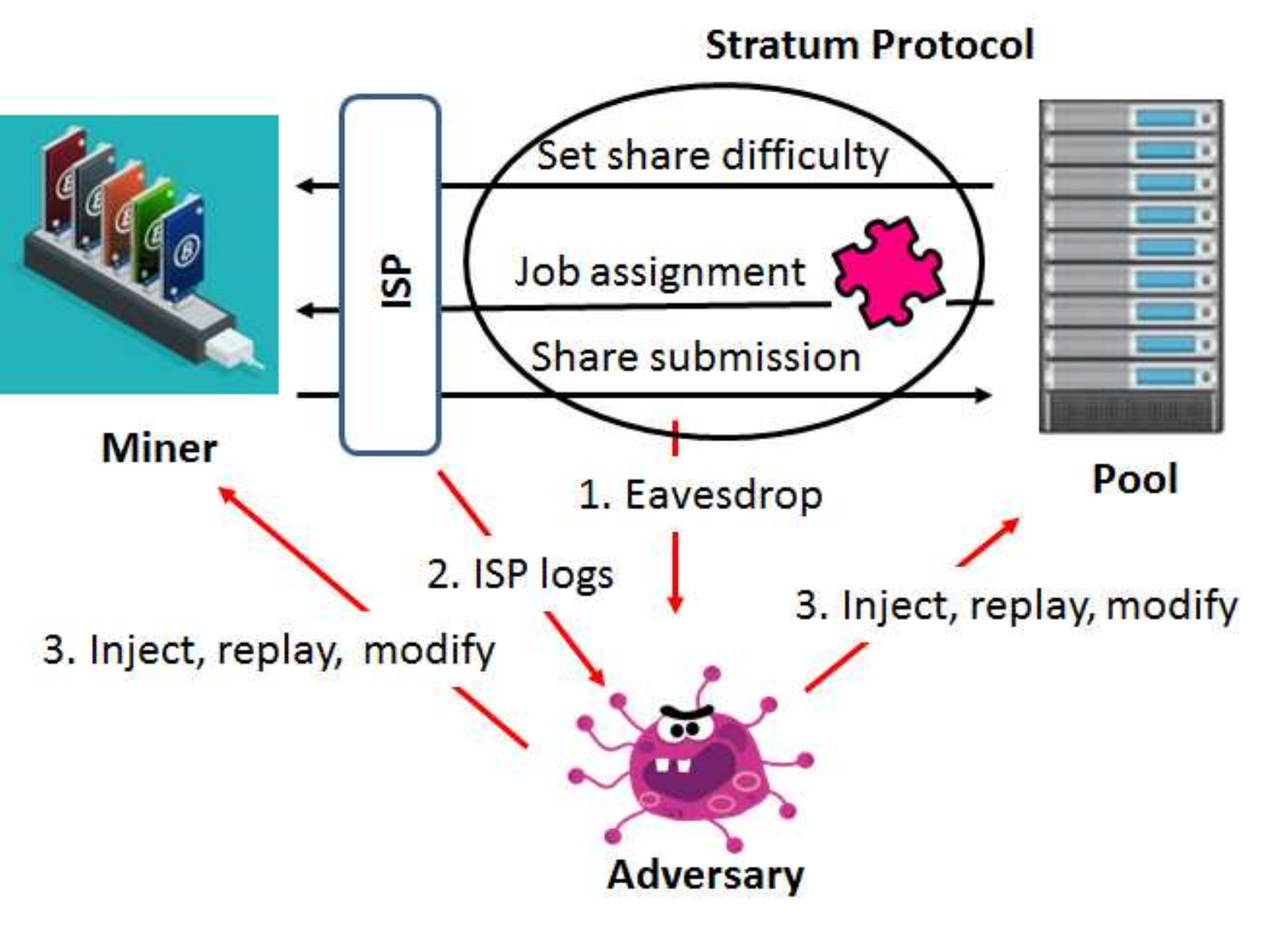}
\caption{Model of system that consists of a pool, miners, and adversary.  The
pool and the miners communicate over the Stratum protocol, to assign jobs and
submit results (shares).  The adversary can eavesdrop, recover ISP logs, inject
and modify the Stratum communications of victim miners.}
\label{fig:system}
\vspace{-15pt}
\end{figure}


We show that the lack of cryptographic protection of communications has made
the Stratum protocol vulnerable to several exploitation possibilities, see
Figure~\ref{fig:system}. An attacker able to observe the traffic between a
miner and a pool server can accurately infer the earnings of the miner. We show
that this result holds even if the attacker has only limited access to the
transmitted packets, e.g., metadata stored in ISP logs.  In addition, we show
that active attackers, able to interfere with the Stratum traffic of miners,
may steal computational resources and bring forth financial loss to their
victims.

These attacks, especially given the wide adoption of the Stratum protocol, show
that Bitcoin and altcoin solutions fail to ensure the monetary privacy and
security of the vital miner community. Furthermore, the attacks reveal that
even an exhaustive use of encryption will fail to ensure miner privacy,
as access to only the timestamp of mining protocol traffic can enable an
attacker to predict the payouts of a victim miner. In addition, the significant
overhead of encryption makes such a solution unappealing to pools, that need to
handle mining traffic from thousands of miners simultaneously, e.g., more than
16,000 for the F2Pool pool~\cite{MinerCount,Corti,F2PoolShare}. In
$\S$~\ref{sec:eval:bedrock} we show that complete encryption of all Stratum
traffic imposes a daily overhead of 1.36 hours on a pool server handling 16,000
miners, while TLS imposes a daily overhead of 1.01 hours.

\newmaterial{
Furthermore, Tor does not address the above vulnerabilities. In fact, sending
Stratum traffic over Tor would enable an adversary to launch the ISP Log attack
not only from the same network with the victim, but also from adversary
controlled Tor exit nodes, that can inspect the cleartext Stratum traffic to
the destination.  Also, Bitcoin over Tor has been shown to be vulnerable to
several attacks~\cite{BP15}, and, even without an adversary, Tor may introduce
delays that can lead to miners losing blocks.}

\noindent
{\bf Our Contributions}.
In this paper we introduce the following contributions:

\begin{compactitem}



\item
{\bf Passive attacks}.
We show that F2Pool's Stratum implementation leaks sensitive miner information
not only through cleartext communications but also indirectly, through hashrate
dependent inter-packet timing.  We introduce {\it StraTap} and {\it ISP Log},
passive attacks where adversaries accurately infer the earnings of victim
miners, given either captured transmissions of those miners, or only their log
metadata.

\item
{\bf Active attack}.
We propose \textit{BiteCoin}, a payout hijack attack that enables an adversary
able to access the communications of a victim miner, to steal its resources and
mining payouts.  To implement BiteCoin, we have developed {\it WireGhost}, a
TCP hijacking tool that surreptitiously modifies TCP packets, without imposing
disconnections or session resets. 

\item
{\bf Bedrock}.
We develop {\it Bedrock}, a Stratum extension that addresses the proposed
attacks.  Bedrock seeks to assuage the efficiency concerns of Bitcoin, by
imposing minimal modifications and encryption overhead to the Stratum protocol.
We introduce the concept of {\it mining cookies}, secret values that miners
need to include in the computed puzzles. Mining cookies prevent both passive
and active attacks on share submission packets, without encrypting the vast
majority of the pool communications.

\item
{\bf Results}.
We have collected 138MB of Stratum traffic traces from mining equipment in the
US and Venezuela, and release it for public use~\cite{StratumData}. We have
implemented the developed attacks and report results from their deployment on
AntMiner mining equipment.  We show that StraTap and ISPLog achieve low payout
prediction errors, and that BiteCoin can efficiently hijack job assignments and
share submissions from a victim miner. We show that Bedrock prevents these
attacks, and is resilient to active attacks even when the adversary breaks its
crypto tools. Bedrock imposes a 12.03s daily overhead on a single pool server
that handles 16,000 miners simultaneously.

\end{compactitem}

\noindent
The attacks and defenses introduced in this paper apply to the Stratum
protocol, thus to most of the large mining
pools~\cite{f2pool_help,AntPool,GHash.io,SlushPool,BTCC}. These attacks work
even on miners that are behind a NAT, or that are firewalled.  Furthermore,
while we focus our experiments on the popular AntMiner mining equipment, our
attacks are general and apply to other manufacturers as well. \newmaterial{We
have notified F2Pool about these vulnerabilities.}

\vspace{-5pt}

\section{Model and Background}
\label{sec:model}


The Bitcoin mining ecosystem consists of miners and pools, see
Figure~\ref{fig:system}. The communication between pools and miners takes place
almost exclusively over the Stratum, that we study in the next section.  The
main task of Bitcoin miners (or mining nodes) is to permanently insert new
consistent data into the network. Miners collect transaction data from other
nodes, validate it and insert it in a structure called \textit{block}. The
miners need to solve a cryptographic puzzle based on the block, before the
block is permanently inserted into the storing structure of the network, called
\textit{the blockchain}.

\subsection{Mining Pools}
\label{sec:model:pool}

\newmaterial{
As specialized, more powerful miners were designed, and the difficulty of
mining blocks increased, it became increasingly difficult for individual, {\it
solo} miners to successfully mine and receive timely payouts for their work.
The concept of {\it mining pools} has emerged in order to address this problem:
enable miners to combine their resources, then split the reward according to
the amount of work they have performed. Pools and miners form a master/slave
paradigm, where pools parallelize the mining work among multiple miners. Thus,
instead of randomly receiving a large reward only once several years, pooled
miners receive smaller rewards, on a regular basis (e.g., once per day).}

\newmaterial{
Popular mining pools today interact with thousands of miners. For instance,
F2Pool has more than 16,000 miners~\cite{MinerCount,Corti,F2PoolShare}. As per
the Bitcoin specification, about 150 blocks are to be mined every day, i.e., 1
block every 10 minutes. Since each block is, at the time of writing, worth 12.5
BTC , at an exchange rate of $\approx$ \$1,100 per BTC, about 2 million dollars
are distributed each day.  The larger mining pool servers control about 20\% of
these earnings~\cite{PoolDistribution}.}

\newmaterial{
The mining pool rewards a {\it share} to the miner who reports a proof of a
unit of solved work.  Pools often offer a {\it variable share difficulty}
feature: adaptively assign the share target to miners, according to their
computation ability. Pools perform this in order to ensure that (i) the
assigned work is not too difficult, thus enable miners to prove computation
progress and gain regular payouts, and that (ii) the work is not too easy, thus
reduce the overhead on the pool, to process shares submitted by thousands of
miners.} 

\noindent
{\bf Hashrate to BTC conversion}.
The pool will reward the miner according to the number of shares submitted and
accepted.  Each pool has a different payout policy. One of the most popular
policies is ``Pay Per Share'' (PPS) which usually has a pool fee attached to
it. Since the final calculation for share payment is dependent on the actual
mining activity of the pool, it is difficult to provide an exact estimate of
each share payout.  However, most pools publish a hashrate to BTC (Bitcoin)
rate of conversion for miner verification purposes. At the time of writing this
paper, the conversion rate on the F2Pool was 0.00246248 BTC per
TH/s~\cite{f2pool_help}.

\subsection{The Coinbase Transaction}
\label{sec:model:coinbase}

\begin{figure}[t!]
\centering
\includegraphics[width=0.49\textwidth]{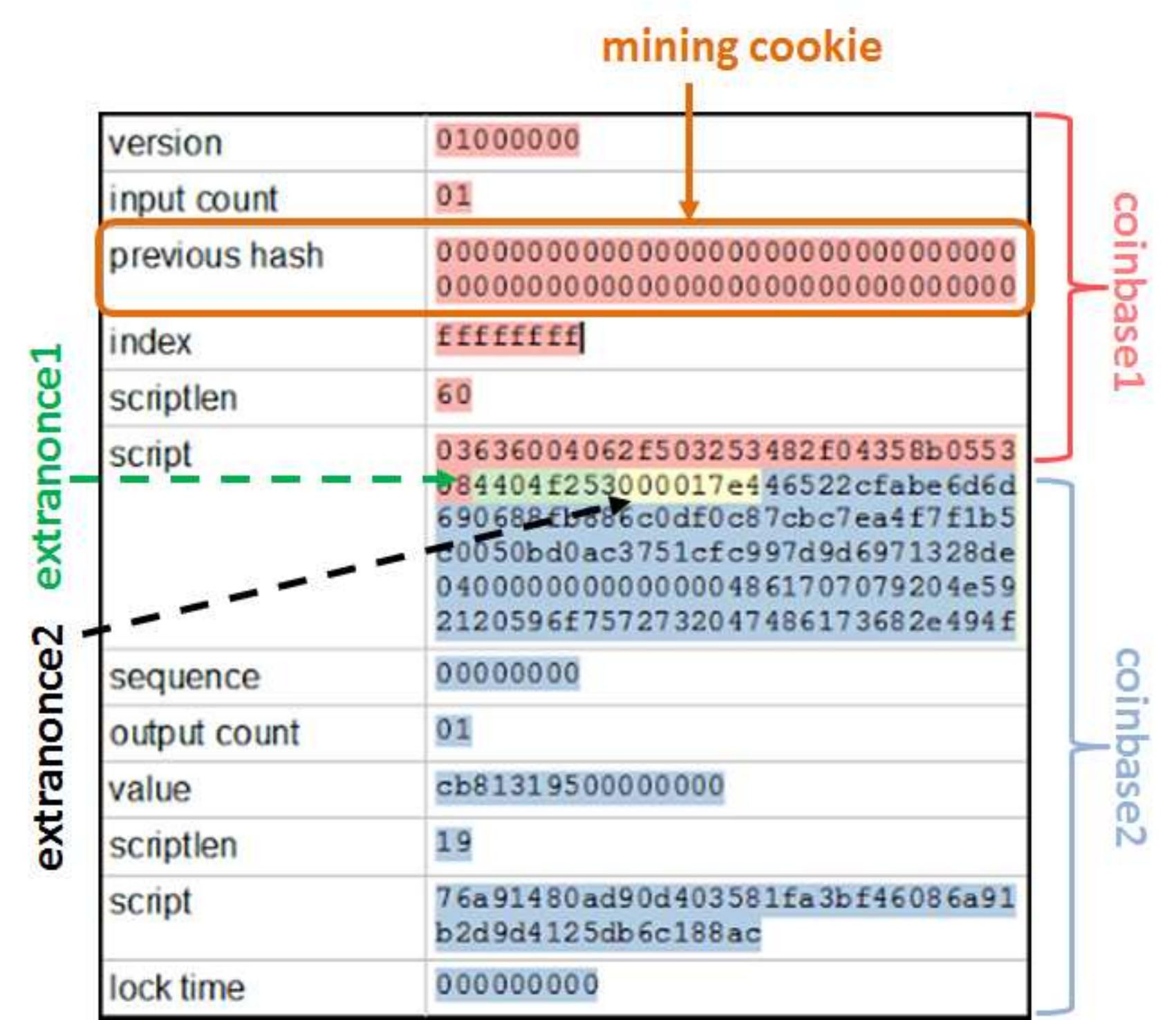}
\caption{Coinbase transaction format. The fields in the table are the
underlying Bitcoin transaction fields. Stratum overlays the coinbase fields
($coinbase1$, $extranonce1$, $extranonce2$, $coinbase2$) on top of the Bitcoin
coinbase fields. Bedrock (see $\S$~\ref{sec:bedrock}) introduces the
{\it mining cookie} concept, whose value will overwrite the currently
unused ``previous hash'' field within $coinbase1$.}
\label{fig:coinbase}
\vspace{-15pt}
\end{figure}

As mentioned above, each block collects a set of previous transactions in the
Bitcoin network. The first such transaction is special, called the {\it
coinbase} transaction (see Figure~\ref{fig:coinbase}): it specifies that the
pool will receive the value of this block (currently 12.5 Bitcoins) when this
block is mined. The Stratum coinbase consists of 4 fields, {\it coinbase1},
{\it extranonce1}, {\it extranonce2}, {\it coinbase2}, overlayed on top of the
Bitcoin coinbase information. $coinbase1$ covers the first 5 fields  of the
input transaction (version, input count, previous transaction hash, previous
transaction index and input scriptlen) and part of the script, in the original
coinbase transaction specification. Except for the version number, these
parameters are meaningless to all clients and pools since the coinbase
transaction does not have an input transaction.
%

The $extranonce1$ and $extranonce2$ fields are also overlayed on the unused
script.  $extranonce1$ needs to be unique (pseudo-random) per stratum
connection.  $extranonce2$ is used in the mining puzzle and needs to be
incremented by the miner once the $nonce$ parameter is exhausted (see next
paragraph). The rest of the coinbase transaction is packed in the $coinbase2$
parameter.

Bedrock, our secure Stratum extension, leverages the unused ``previous
transaction hash'' field (Figure~\ref{fig:coinbase}), to include the value of
the mining cookie, see $\S$~\ref{sec:bedrock}.

\begin{figure}[t]
\centering
\includegraphics[width=0.47\textwidth]{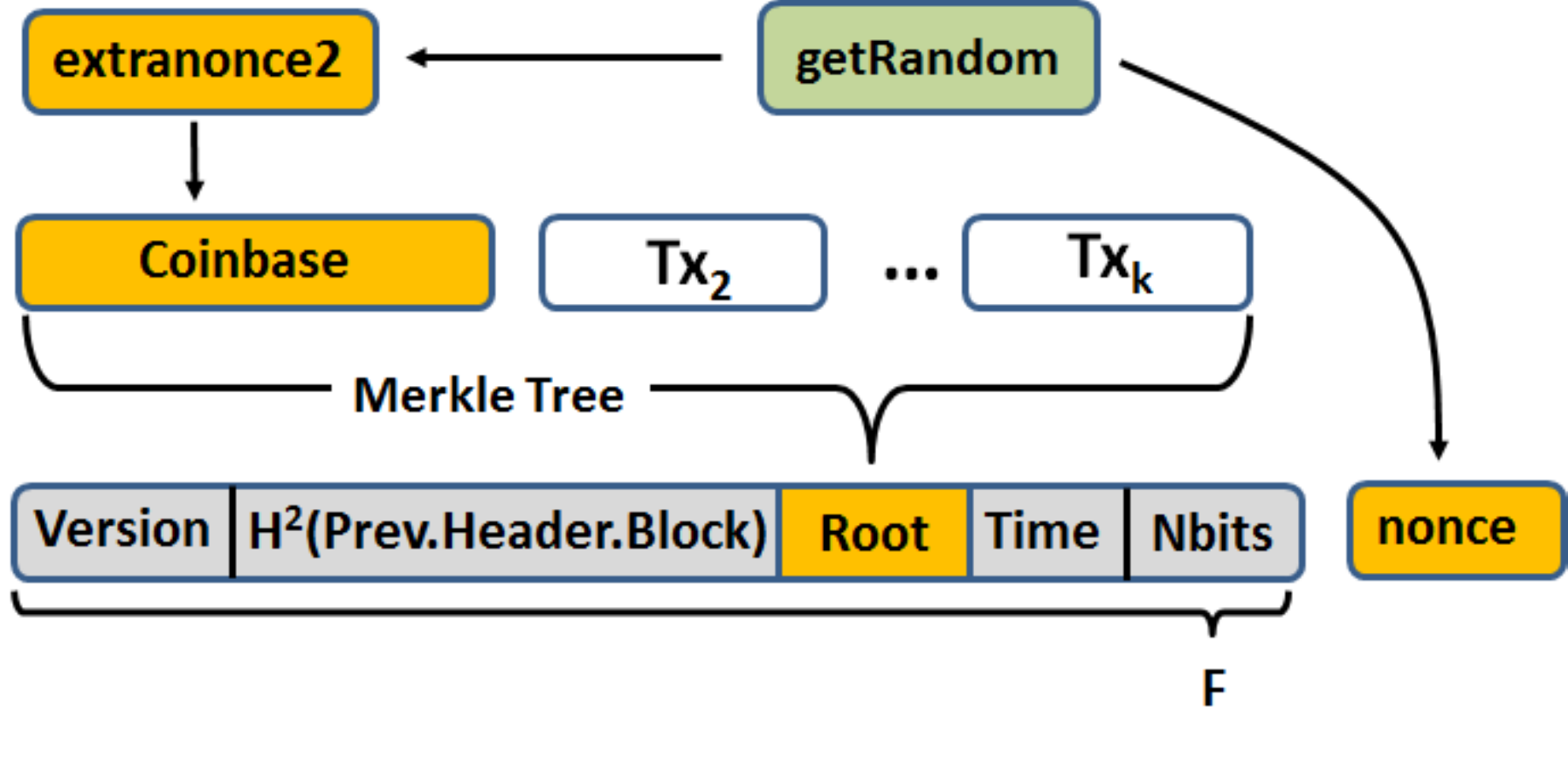}
\vspace{-15pt}
\caption{Bitcoin block puzzle. The root of the Merkle tree built over
the coinbase and the mined transactions is the third field of the block
puzzle. The miner iterates over the $nonce$ (last field) and over the
extranonce2 value, part of the coinbase transaction.}
\label{fig:puzzle}
\vspace{-10pt}
\end{figure}

\subsection{The Bitcoin Puzzle}
\label{sec:model:puzzle}

The goal of the mining process is to make it difficult for a minority of
malicious nodes to insert bogus data inside invalid blocks. It achieves this by
transforming each block (collection of transactions in the Bitcoin network)
into a cryptographic ``puzzle''. The puzzle is designed such that the
probability of finding a solution by a mining node is proportional to its
computational power.  A Bitcoin puzzle consists of a $target$ value and a
tuple F = (block version number || hash of previous block || RMT ||
timestamp || Nbits ), || denotes concatenation, see
Figure~\ref{fig:puzzle} for an illustration.  


Specifically, F contains the block version number, the hash of the previous
block in the blockchain, the root of a Merkle tree (RMT) described next, a
timestamp, and the final target value in the form of the number of leading bits
that need to be 0 for the block to be considered ``mined''. The
Merkle tree is built over the transactions that are being mined into the
current block, including the coinbase transaction, see Figure~\ref{fig:puzzle}.
Given the F value, and the above mentioned $target$, the miner iterates
over the $nonce$ and $extranonce2$ (see coinbase transaction) values until it
identifies a pair such that
\vspace{-5pt}

\begin{equation}
H^2(nonce || F) < target
\label{eq:puzzle}
\end{equation}

\noindent
where $H^2$ denotes the double (SHA-256) hash. The block is said to be
``mined'' when $H^2(nonce || F)$ is less than the target corresponding to the
above Nbits value.

\noindent
{\bf The target and the difficulty}.
While the $Nbits$ value specifies when the block is mined, pools set the above
$target$ parameter to a larger value (fewer leading bits 0) to enable miners
to prove and be rewarded for progress. The {\it target of difficulty 1},
denoted $target\_1$, is defined by pools as the number $2^{224} - 1$, i.e. a
256 bit number with 32 bits of leading zeros followed by 224 bits of ones.  The
\textit{difficulty} value is a measure of how hard it is to solve a puzzle for
a given $target$ value.  Accordingly, the relationship between the $difficulty$
and $target$ values is given by the formula:
\vspace{-5pt}

\begin{equation}
difficulty = \frac {target\_1}{target} = \frac{2^{224} - 1}{target}
\label{eq:difficulty}
\end{equation}

\vspace{-5pt}

\subsection{\newmaterial{Stratum}}
\label{sec:stratum}

\begin{figure}[t!]
\centering
\includegraphics[width=0.49\textwidth]{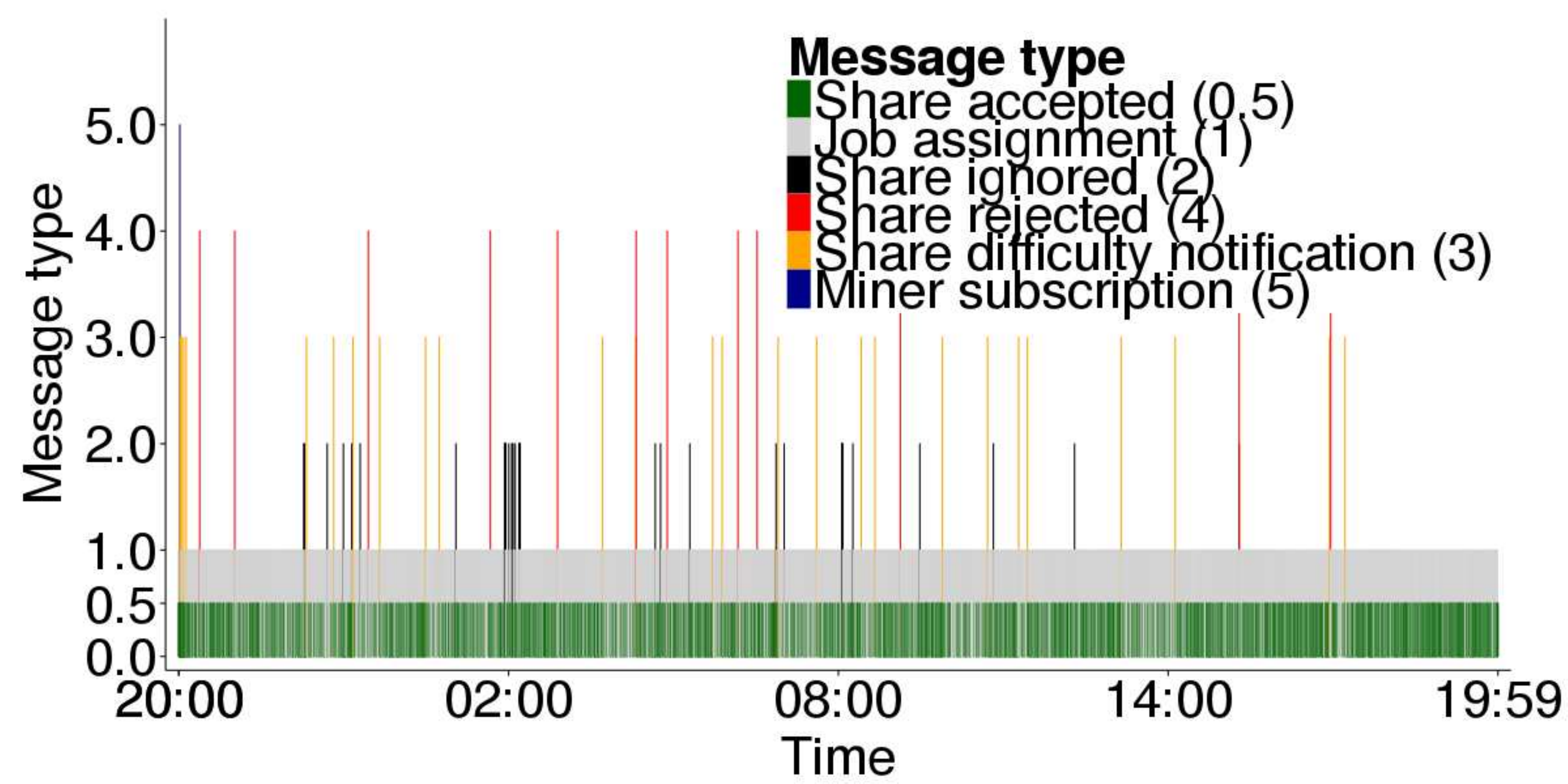}
\caption{Stratum protocol timeline over 24 hours captured between an AntMiner
S7 and the F2Pool pool. While we observe several difficulty change packets
(orange bars) throughout the day, most are concentrated after the initial
subscription protocol (tall blue bar). The majority of share submissions
are accepted (green bars); only a few are rejected (red bars) or ignored
(black bars).}
\label{fig:timeline:general}
\vspace{-15pt}
\end{figure}

Stratum is a clear text communication protocol between the pool and the
miners~\cite{stratum_proto}, built over TCP/IP and using the JSON-RPC format.
\newmaterial{
The official Stratum protocol documentation is not detailed and is often
outdated~\cite{stratum_proto}.  In this section we describe the
F2Pool~\cite{f2pool_help} implementation of the Stratum protocol, as observed
from Stratum packets we captured over 13 days between an AntMiner S7 device and
the F2Pool mining pool (see $\S$~\ref{sec:implementation:passive}).}
Figure~\ref{fig:timeline:general} illustrates the timeline of captured Stratum
protocol packets over a 24 hour interval. The ability to capture, understand,
modify and inject these messages into a communication stream will be
instrumental for the passive and active attacks described in
$\S$~\ref{sec:attacks:passive} and $\S$~\ref{sec:attacks:active}.


%
%

\noindent
{\bf Miner subscription}.
\newmaterial{
To register with the pool, the miner first subscribes through a
\textbf{connection subscription} message
\vspace{-5pt}
\[
{\tt mining.subscribe,\ params},
\]
that describes the miner capabilities. The server responds with a
\textbf{subscription response} message,
\vspace{-5pt}
\[
{\tt result, \{methods\},\ extranonce1,\ extranonce2.size},
\]
where the first field is a list method names used by the server pool, the
second field (see $\S$~\ref{sec:model:puzzle}), should be random and unique per
connection, but F2Pool sets to constant ``$\backslash x30\backslash x30$'', and
the third is the size of the {\tt extranonce2} (4B in F2Pool).}

\begin{figure}[t]
\centering
\includegraphics[width=0.49\textwidth]{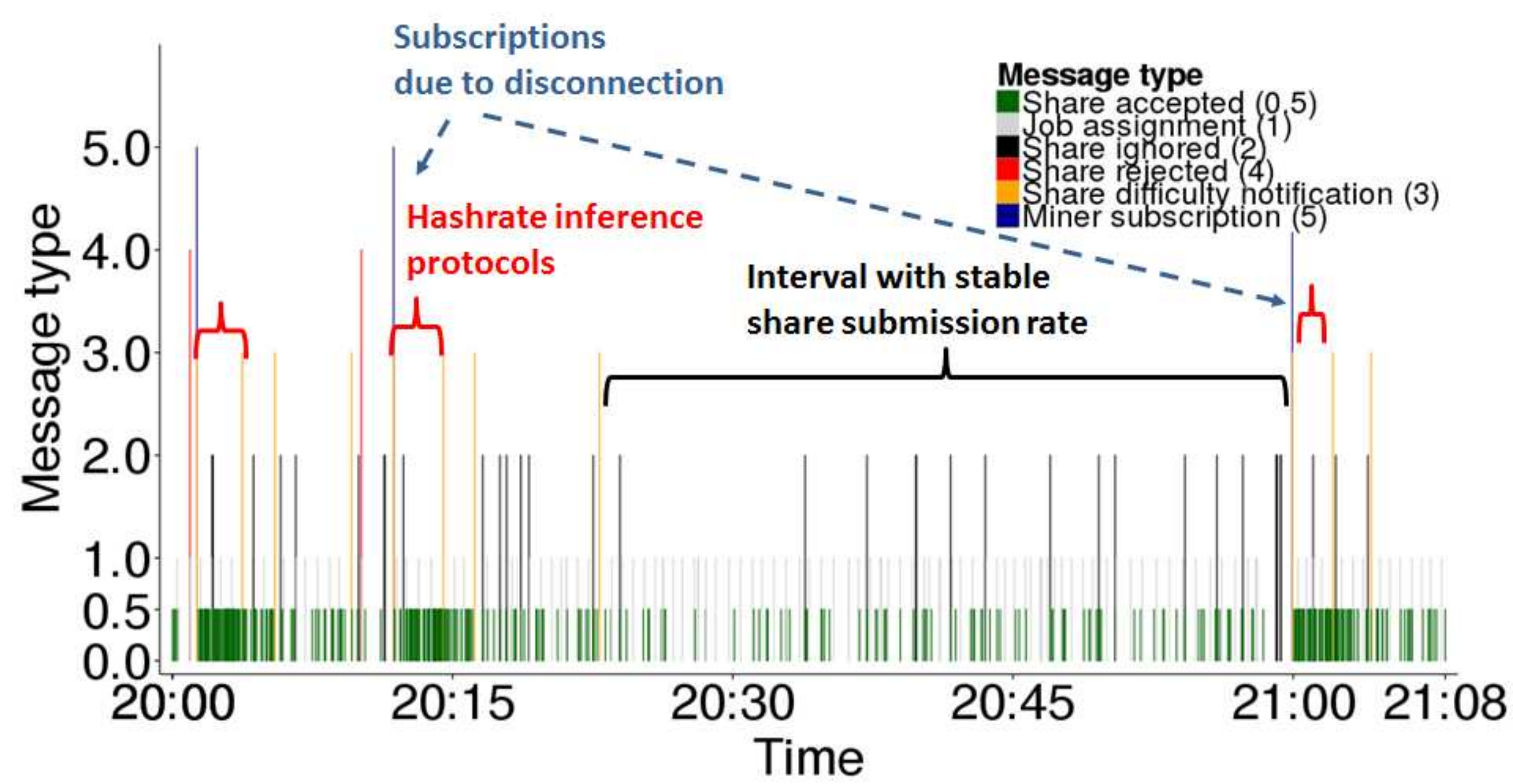}
\caption{Timeline of miner operating at 250MHz. We emphasize the effect
of successive share difficulty notification messages: the miner's share submission
rate slows down. We also point the multiple miner subscribe and authorization
procedures (tall blue bars) due to repeated miner disconnections.}
\label{fig:timeline:rate}
\vspace{-15pt}
\end{figure}

\noindent
\textbf{Miner authorization}.
\newmaterial{
Following the subscription exchange, Stratum authenticates the miner with the
pool, through a \textbf{miner authorization request} message
\vspace{-5pt}
\[
{\tt mining.authorize,\ account.minerID,\ password},
\]
whose first field, the username, consists of two fields, that enable a user to
register multiple miners with the same account. While the password
field is transferred in cleartext, it is currently ignored by pools.
%
%
The pool responds with a \textbf{status result} that notifies the miner of the
result of the authorization request.} In Figure~\ref{fig:timeline:general}
and~\ref{fig:timeline:rate}, the miner subscription and authorization messages
are shown as a single blue bar (the tallest), seen at the beginning of the
interval and each time the miner reconnects to the pool, e.g., after an
Internet disconnection or power outage.

\noindent
{\bf Share difficulty notification}.
\newmaterial{
Following a successful authorization, the pool sends a \textbf{difficulty
notification} message to the miner
\vspace{-5pt}
\[
{\tt set.difficulty,\ difficulty},
\]
that specifies the minimum share difficulty that the server will be willing to
accept.}
Figure~\ref{fig:timeline:general} and~\ref{fig:timeline:rate} show
the difficulty notification messages as orange bars. They can occur
throughout the day as the pool seeks to adjust the miner's rate of share
submissions.
%

\noindent
{\bf Job assignment}.
\newmaterial{
The pool assigns jobs (puzzles) to the miner through \textbf{mining job
notification} messages
\vspace{-5pt}
\[
{\tt mining.notify,\ job\_id,\ params,\ clean\_jobs}
\]
that specify the puzzle parameters, i.e., the fields of the $F$ value in
Equation~\ref{eq:puzzle} (see $\S$~\ref{sec:model:puzzle}),
and a boolean that indicates if the miner should drop all previous jobs and
work exclusively on the one specified by this message.}


\noindent
\textbf{Share submission}.
\newmaterial{
Once the miner finds a solution that satisfies Equation~\ref{eq:puzzle}, it
sends a {\bf share submission} message to the pool for verification and credit:
\[
{\tt mining.submit,\ account.minerID,\ job\_id,\ time,}
\]
\vspace{-20pt}
\[
{\tt nonce,\ extranonce2}
\]
that specifies the miner's username, the job id received in the previous mining
job notification, and the parameters of the puzzle solution: the $nonce$ and
$extranonce2$ parameters, see $\S$~\ref{sec:model:puzzle}.  The pool uses these
values to reconstruct the $F$ value (see $\S$~\ref{sec:model:puzzle}), and
verifies that Equation~\ref{eq:puzzle} is satisfied.}

The pool responds with a status result message, illustrated in
Figure~\ref{fig:timeline:general}: green bars denote accepted shares, red bars
denote rejected shares, and black bars denote ignored shares. Shares can be
rejected due to stale work, i.e., being submitted too late.
The miner continues to mine current jobs until it receives a job message from
the pool that requires it to invalidate previous jobs (see the ``clean jobs''
flag in the job assignment message).

\vspace{-5pt}

\section{Adversary Model}
\label{sec:model:adversary}

We consider adversaries that can launch both passive and active attacks against
the Bitcoin network, see Figure~\ref{fig:system}. We assume that the pool and
the miner are honest. However, adversaries can target the communications of
specific, victim miners. Adversaries can own mining equipment, can eavesdrop
and interfere with existing communications, and may even obtain data from ISPs.
We now detail each of these adversarial capabilities.

\noindent
{\bf Eavesdropping capabilities}.
We consider first an adversary who can access the entire communication of a
victim miner. Such adversaries include over-controlling governments, or
attackers who gain control to equipment on the same LAN as the victim. We
assume that such an adversary can capture and inspect all the packets sent and
received by the victim miner.

\noindent
{\bf ISP log capabilities}.
We also consider adversaries with access to ISP logs, that include entries for
the communications of the miners in the ISP's subnet and the pool. This
capability is more restrictive than the eavesdropping capability, in terms of
the data that can be extracted from the miner-to-pool communications. This is
because ISP logs usually contain only metadata~\cite{dataRetention}, in order
to comply with law enforcement requests~\cite{desimmone2010pitting}. However,
these capabilities may enable the adversary to target more victims (i.e., all
the miners whose traffic was logged).

Potential perpetrators include insiders (e.g., ISP employees) and government
organizations that can subpoena the logs.  While law enforcement insiders have
been shown to abuse collected data~\cite{nsaSpy}, agencies have also been
hacked in the past. Stolen data may then be sold, auctioned, or even made
public, thus becoming accessible to a broader range of adversaries.

We assume that this adversary has access to packet metadata that includes
timestamps, source and destination IPs and ports, and connection flags. As we
show in $\S$~\ref{sec:attacks:passive:isp}, these values enable the attacker to
identify mining traffic via well known pool IP/port pairs, and identify the
start of mining sessions.

\noindent
{\bf Active attack capabilities}.
We further consider an adversary that can modify the communication stream
between the server pool and a mining client. Potential such adversaries include
attackers that are on the same network as the victim miner, rogue employees at
an intermediate ISP, or a government backed agency. In
$\S$~\ref{sec:attacks:active}, we show that such an adversary can add jobs and
replace submitted shares.  While this may allow for trivial denial of service
attacks as well, in this paper we do not consider DoS attacks. We note that DoS
techniques exist with lesser technical requirements~\cite{caooff,qian2012off}.

\vspace{-5pt}

\subsection{Relevance of Attacks}

\newmaterial{
In the following section we introduce passive and active attacks against miners
that use the Stratum protocol, whose goal ranges from inferring to stealing the
payouts of the victim miners. Inferring the payouts of miners exposes their
owners to a suite of attacks. Adversaries can hack the computers or accounts of
miner owners in order to steal their
payouts~\cite{BitcoinTheft,ListBitcoinTheft}.  Passive attacks can also enable
the adversary to identify miners worthy of being targeted with resource
hijacking attacks, e.g., Bitecoin, see $\S$~\ref{sec:attacks:active}.  In
addition, this knowledge enables adversaries to target miner owners for
equipment theft, arson~\cite{Arson}, kidnapping~\cite{williams2011kidnappings},
and, in countries where Bitcoin is illegal~\cite{WikiIllegal,BitcoinIllegal},
for extortion and prosecution~\cite{VenezuelaBitcoin}.}

\vspace{-5pt}

\section{Passive Attacks}
\label{sec:attacks:passive}

\vspace{-5pt}

In this section we show that an attacker that observes even partial traffic
of a victim miner, can infer the payouts received by the miner. We introduce
two passive attacks, that make different assumptions on the adversary's
capabilities. First, the StraTap attack assumes an adversary able to capture
and inspect entire packets transmitted between a pool and a victim miner.
Second, the ``ISP Log'' attack assumes an adversary that is able to inspect
only packet metadata, i.e., IP addresses, port numbers, and connection flags.
In the following we detail each attack.

\vspace{-5pt}

\subsection{The StraTap Attack}
\label{sec:attacks:passive:full}

\vspace{-5pt}

We consider first an eavesdropping adversary, see
$\S$~\ref{sec:model:adversary}.  Given access to all the packets sent and
received by the victim miner, the attacker counts the share submission messages
along with their associated difficulty (as described in
$\S$~\ref{sec:stratum}). The attacker uses this data to estimate the hashrate
of the victim miner.

Specifically, the probability of randomly finding a hash with the appropriate
difficulty is given by the ratio between the target (i.e.  the number of hashes
with the appropriate number of leading zeros according to the assigned job),
and the total number of possible hashes. Hence, the probability of a miner
finding a share \newmaterial{with a single hash} is $p = \frac{target}{ 2^{256}
- 1}$.  The expected number of hashes, $E$, that the miner needs to calculate
before finding a valid share is then $1/p$. Then, we derive the following for
$E$:

\begin{equation}
\begin{split}
E &= \frac{ 2^{256} - 1}{target\_1}\times \frac{target\_1}{target} = 
\\
&= \frac{ 2^{256} - 1}{2^{224} - 1} \times difficulty \approx difficulty \times 2^{32}
\end{split}
\label{eq:e}
\end{equation}

\noindent
The second equality follows from Equation~\ref{eq:difficulty}, and the fact
that the target of difficulty 1, $target\_1$ is $2^{224} -1$, see
$\S$~\ref{sec:model:puzzle}.
If we divide Equation~\ref{eq:e} by the hashrate of the miner, we obtain a
formula that allows us to compute the expected time to find a share at a given
difficulty:

\begin{equation}
\begin{split}\
time &= \frac{E}{hashrate} = \frac{difficulty \times 2^{32}}{hashrate}
\end{split}
\label{eq:expected_time_to_share}
\end{equation}

\noindent
Thus, $hashrate = difficulty \times \frac{2^{32}}{time}$.
The attacker obtains the $difficulty$ value by inspecting the \textit{share
difficulty notification} messages (see $\S$~\ref{sec:stratum}).  In
addition, the attacker estimates the $time$ value as the ratio of the length of
time between consecutive share difficulty notification messages (orange bars in
Figure~\ref{fig:timeline:general}) and the number of shares submitted and
accepted during that interval: \[ \overline{time} = \frac{\text{total time
between difficulty changes}}{\text{number of submitted (and approved) shares}}
\] The attacker obtains the accepted share count by inspecting the share
submission messages and their corresponding status results, see
$\S$~\ref{sec:stratum}.

Given the inferred $hashrate$ of the miner, the attacker uses the hashrate to
payout conversion (see the corresponding paragraph in $\S$~\ref{sec:stratum})
to predict the amount of Bitcoins received by the victim. In
$\S$~\ref{sec:eval:passive:stratap} we experimentally evaluate the accuracy
prediction of the StraTap attack.

\subsection{The ISP Log Attack}
\label{sec:attacks:passive:isp}

We now consider an attacker with ISP log data capabilities, see
$\S$~\ref{sec:model:adversary}.
%
%
The ISP Log attack proceeds as follows. First, the attacker identifies the
beginning of the connection between the victim miner and the pool. This is the
time when a 3-way handshake connection is established, whose first step is a
``connection subscription'' message as described in $\S$~\ref{sec:stratum}.
Then, the attacker predicts the hashrate of the miner based on statistics over
the inter-packet times logged for the miner.

%

In early experiments we have observed that predictors that use statistics over
long time intervals are inaccurate.  To address this problem, we have
identified and exploited a vulnerability of the Stratum protocol. Specifically,
we observed that the first share difficulty notification message (see
$\S$~\ref{sec:stratum}) following a successful miner subscription, sets the
difficulty to the minimum value acceptable by the pool (e.g., 1024 for F2Pool).
In addition, in $\S$~\ref{sec:eval:passive:isp} we show that the pool sends its
second share difficulty notification after approximately 50 share submission
messages (for difficulty 1024) received from the miner.

\begin{figure}[t]
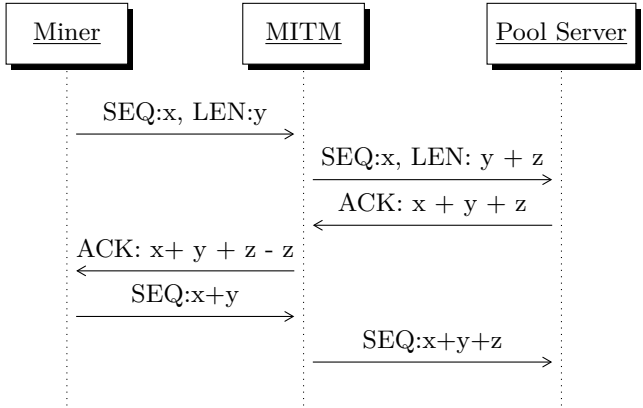

\begin{sequencediagram}
  \newinst[0.01]{a}{Miner}
  \newinst[1.5]{m}{MITM}
  \newinst[1.6]{b}{Pool Server}

  \mess{a}{SEQ:x, LEN:y}{m}
  \mess{m}{SEQ:x, LEN: y + z }{b}
  \mess{b}{ACK: x + y + z}{m}
  \mess{m}{ACK: x+ y + z - z }{a}
  \mess{a}{SEQ:x+y}{m}
  \mess{m}{SEQ:x+y+z}{b}
\end{sequencediagram}
\caption{WireGhost illustration: TCP hijacking with active re-synchronization,
when the man-in-the-middle (MITM) adversary adds $z$ bytes to an existing
packet originating from the client. WireGhost modifies the sequence numbers, to
hide the difference in packet sizes.}
\label{fig:active:tcp_hijack}
\vspace{-5pt}
\end{figure}

Then, the attacker estimates the time taken by share submissions for jobs of
difficulty 1024, i.e., over the first 50 packets sent by the miner following
its subscription and authorization process.  It then uses the process outlined
in the above StraTap attack to predict the miner's hashrate and payout. The
attacker can repeat this process when observing subsequent 3-way handshake
connection protocols of the victim miner, e.g., when a disconnection occurs, in
order to improve its estimates of the miner hashrate. In
$\S$~\ref{sec:eval:passive:isp} we show that even when the ISP Log attack
performs a single hashrate inference attack per day, its daily miner payout
prediction achieves a mean percentage error of -9.49\%.

\vspace{-10pt}

\section{The BiteCoin Attack}
\label{sec:attacks:active}

\vspace{-5pt}

We consider now an active attacker with the ability to capture and modify the
communication stream between the pool and the victim miner, see
$\S$~\ref{sec:model:adversary}. In the following, we first focus on the
challenge to hijack and maintain the TCP connection between the miner and the
pool, then introduce BiteCoin, an attack that hijacks payments from victim
miners.

\vspace{-5pt}

\subsection{WireGhost: TCP Hijack with Re-Sync}

\vspace{-5pt}

\noindent
{\bf Existing tools}.
Traditional TCP hijacking attack tools seldom consider the need to
preserve the status of the communication parties. For instance, in tools like
Shijack~\cite{toolShijack} and Juggernaut~\cite{toolJuggernaut}, once the TCP
sequence mangling is performed, the generated ack storm is eliminated by
resetting the connection with one of the peers. The tool Hunt~\cite{toolHunt}
does have a re-synchronization functionality but it is limited to the Telnet
protocol and requires victim interaction in the form of a social engineering
attack to be successful. Stratum active attacks require that the original
mining connection is maintained and that the re-synchronization needs to be
done completely unattended. For instance, the $extranonce1$ parameter will be
different for each connection, thus the attacker should not force
disconnections.


\noindent
{\bf WireGhost}.
We have developed WireGhost, a TCP hijack tool that maintains
the status of the hijacked connection, without having to reset the
communication streams.  To address ack storms that would occur due to
communication changes (e.g., packet modification, injection, removal), WireGhost
modifies the TCP sequence of packets according to the payload modification
performed by the attacker, see Figure~\ref{fig:active:tcp_hijack}.
Specifically, if the attacker inserts data into the TCP payload (including
injecting new packets), WireGhost subtracts the appropriate number of bytes
from the pool server's ack sequence for all the packets that follow the
modified (or inserted) one. It then adds the same amount of bytes 
to the sequence number for all the following packets originating from the
client. WireGhost performs the opposite mathematical operations when the
attacker removes data from the TCP payload. 

\begin{figure}[t]
\includegraphics[width=0.49\textwidth]{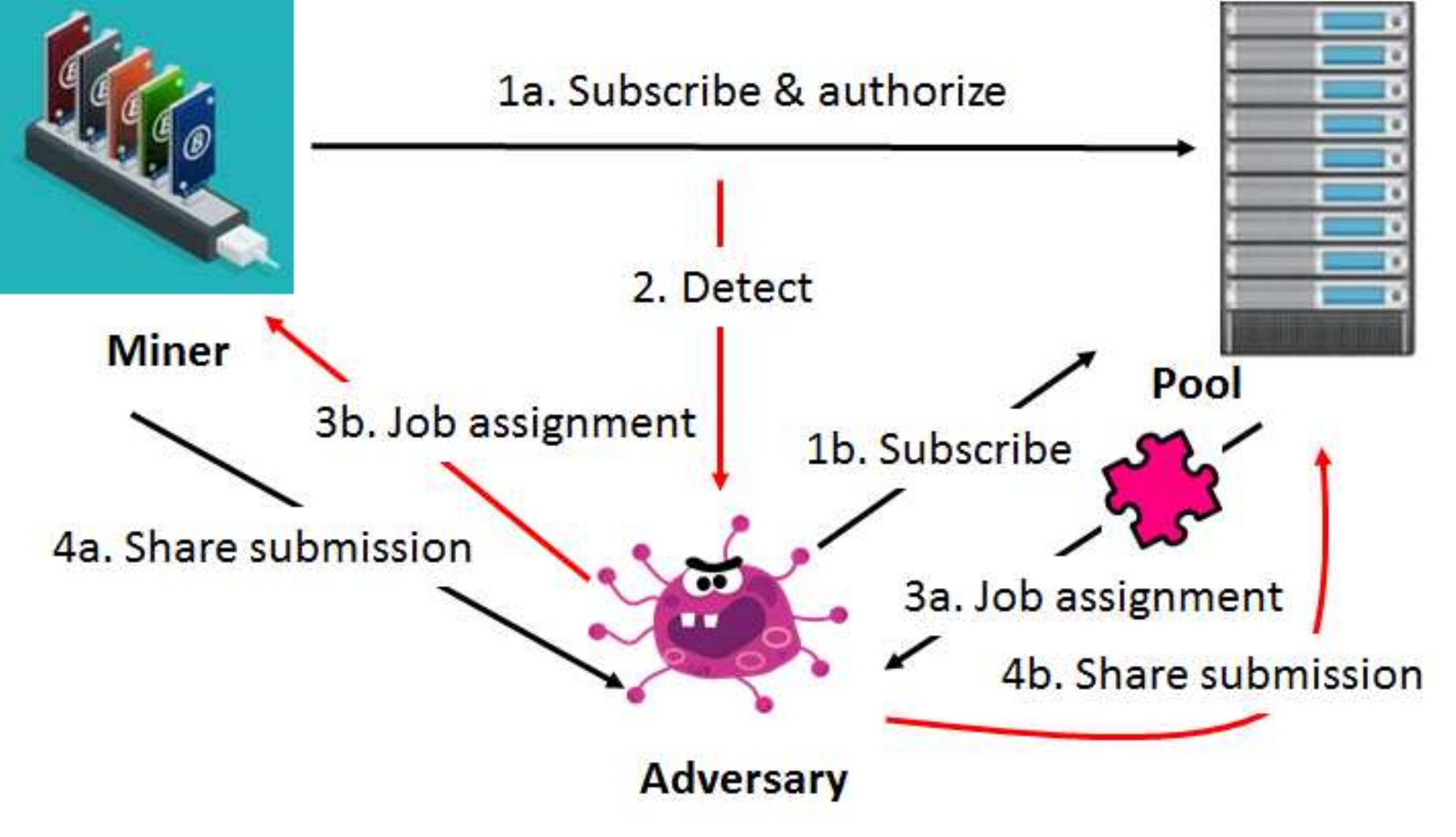}
\caption{BiteCoin attack illustration. The attacker, a subscribed miner,
forwards job assignments received from the pool to the victim miner. It
then hijacks the victim's share submissions and sends them as its own
to the pool, to get the credit.}
\label{fig:bitecoin}
\vspace{-15pt}
\end{figure}

\vspace{-5pt}

\subsection{BiteCoin}

\vspace{-5pt}


We have used WireGhost to develop BiteCoin, an attack tool that enables an
active adversary to steal CPU cycles and payouts from victim miners.  We
consider an attacker who subscribes a device under his control as a miner to
the pool, see Figure~\ref{fig:bitecoin} for an illustration.

Given access to the communication medium between the victim miner and the pool,
BiteCoin first detects the miner subscription protocol (the 3-way handshake).
It then uses WireGhost to hijack the TCP connection between the miner and the
pool. Then, when the attacker device receives a job assignment message from the
pool, it directly injects it into the TCP connection of the victim miner and
the pool.  The victim miner receives this job assignment packet as if it was
coming from the pool.

Once the victim miner computes a share for this job, it packs it into a share
submission message and sends it over its TCP connection to the pool. BiteCoin
intercepts this share submission packet of the victim, and modifies it by
changing the victim's username to its own. It then sends this modified share
submission over its own TCP connection to the pool.  BiteCoin also sends to the
pool a mangled copy of the victim's original share submission, to ensure that
it is rejected.  In $\S$~\ref{sec:implementation:bitecoin} we detail our
BiteCoin implementation, and in $\S$~\ref{sec:eval:bitecoin} we present results
over its deployment.


\vspace{-10pt}

\section{Bedrock: Secure Stratum}
\label{sec:bedrock}

\vspace{-5pt}

We now study defenses against the proposed attacks. We first describe the
requirements of a private and secure mining protocol, then introduce Bedrock, a
Stratum extension, and discuss its defenses.

\vspace{-5pt}

\subsection{Solution Requirements}
\label{sec:bedrock:requirements}

\vspace{-5pt}

A private and secure Stratum protocol should satisfy the following informal
requirements:

\begin{compactitem}

\item
{\bf Security}.
The solution needs to protect both against the Stratum attacks that we
introduced in $\S$~\ref{sec:attacks:passive} and $\S$~\ref{sec:attacks:active},
and against attacks that target the solution itself.

\item
{\bf Efficiency}.
Encryption of all the Stratum messages is not only inefficient, but also
insecure: in $\S$~\ref{sec:eval:passive:isp} we show that the ISP Log attack
can predict the miner's profits while knowing only the miner's transmission
timestamps.

\item
{\bf Adoptability}.
The solution should introduce minimal modifications to the Stratum protocol,
in order to simplify its adoptability by pools and miners.

\end{compactitem}

\vspace{-5pt}

\subsection{The Solution}
\label{sec:bedrock:solution}

We introduce Bedrock \footnote{In geology, the bedrock is a hard stratum.}, a
secure and efficient extension of the Stratum protocol. Bedrock seeks to
prevent adversaries from inferring the hashrates of miners, and to efficiently
authenticate Stratum messages.

\begin{figure}[h]
\vspace{-15pt}
\renewcommand{\baselinestretch}{0.5}
\begin{minipage}{0.49\textwidth}
\begin{algorithm}[H]
\begin{tabbing}
XXX\=X\=X\=X\=X\=X\= \kill

1.{Implementation\ \mbox{\bf{PoolServer}}}\\
2.\>{\mbox{\bf{generateCookie}}(Miner $M$)\{}\\
3.\>\>{$R_M$ := getRandom(256);}\\
4.\>\>{$C_M$ := $H^2(R_M, M.uname)$;}\\
5.\>\>{$K_M$ := $M.key$;}\\
6.\>\>{store($M.uname$, $K_M$, $R_M$, $target$);}\\
7.\>\>{sendEncrypted($M$, $E_{K_M}(R_M)$);}\\

8.\>{\mbox{\bf{verifyJob}}(Miner $M$,\ $nonce$,\ $extranonce2$)\{}\\
9.\>\>{($K_M$,\ $R_M$,\ $target$) := getMParams($M.uname$);}\\
10.\>\>{$C_M$ := $H^2(R_M, M.uname)$;}\\
11.\>\>{$F$ := computeF($C_M$,\ $extranonce2$);}\\
12.\>\>{\mbox{\bf{if}}\ ($H^2(nonce || F) < target$)}\\
13.\>\>\>{sendToMiner($M$, result, ``accept'');}\\
14.\>\>{\mbox{\bf{else}}\ sendToMiner($M$, result, ``reject'');}\\\\

15.{Implementation\ \mbox{\bf{Miner}}}\\
16.\>{$K_M$ : int[256] \% key\ shared\ with\ pool}\\
17.\>{$C_M$ : int[256] \% mining\ cookie;}\\
18.\>{\mbox{\bf{solvePuzzle}}($target$: int)\{}\\
19.\>\>{\mbox{\bf{do}}}\\
20.\>\>\>{$randPerm$ := newPseudoRandPerm(32);}\\
21.\>\>\>{$extranonce2$ := getRandom(32);}\\
22.\>\>\>{$F$ := computeF($C_M$,\ $extranonce2$);}\\
23.\>\>\>{\mbox{\bf{while}}\ ($randPerm$.isNext())\{}\\
24.\>\>\>\>{$nonce$ := $randPerm$.next();}\\
25.\>\>\>\>{\mbox{\bf{if}}\ ($H^2(nonce || F) < target$)}\\
26.\>\>\>\>\>{sendToPool($uname$, $nonce$, $extranonce2$);}\\
27.\>\>{\mbox{\bf{while}} ($clean\_jobs$ != 1)}
\vspace{-25pt}
\end{tabbing}
\caption{Bedrock pseudo-code for cookie generation and job verification (pool
side), and job solving (miner side).}
\label{alg:bedrock}
\end{algorithm}
\end{minipage}
\normalsize
\vspace{-15pt}
\end{figure}


Bedrock has 3 components, each addressing different Stratum vulnerabilities.
The first component authenticates and obfuscates the job assignment and share
submission messages. The second component secures the share difficulty
notifications, and the third component secures the pool's inference of the
miner's capabilities.  In the following we detail each component. We assume
that the pool shares a unique secret symmetric key $K_M$ with each miner $M$.
The miner and the pool create the key during the first authorization protocol
(see $\S$~\ref{sec:stratum}), e.g., using authenticated Diffie-Hellman).

\vspace{-5pt}

\subsubsection{Mining Cookies}
\label{sec:bedrock:solution:cookies}

\begin{figure}[t]
\centering
\includegraphics[width=0.47\textwidth]{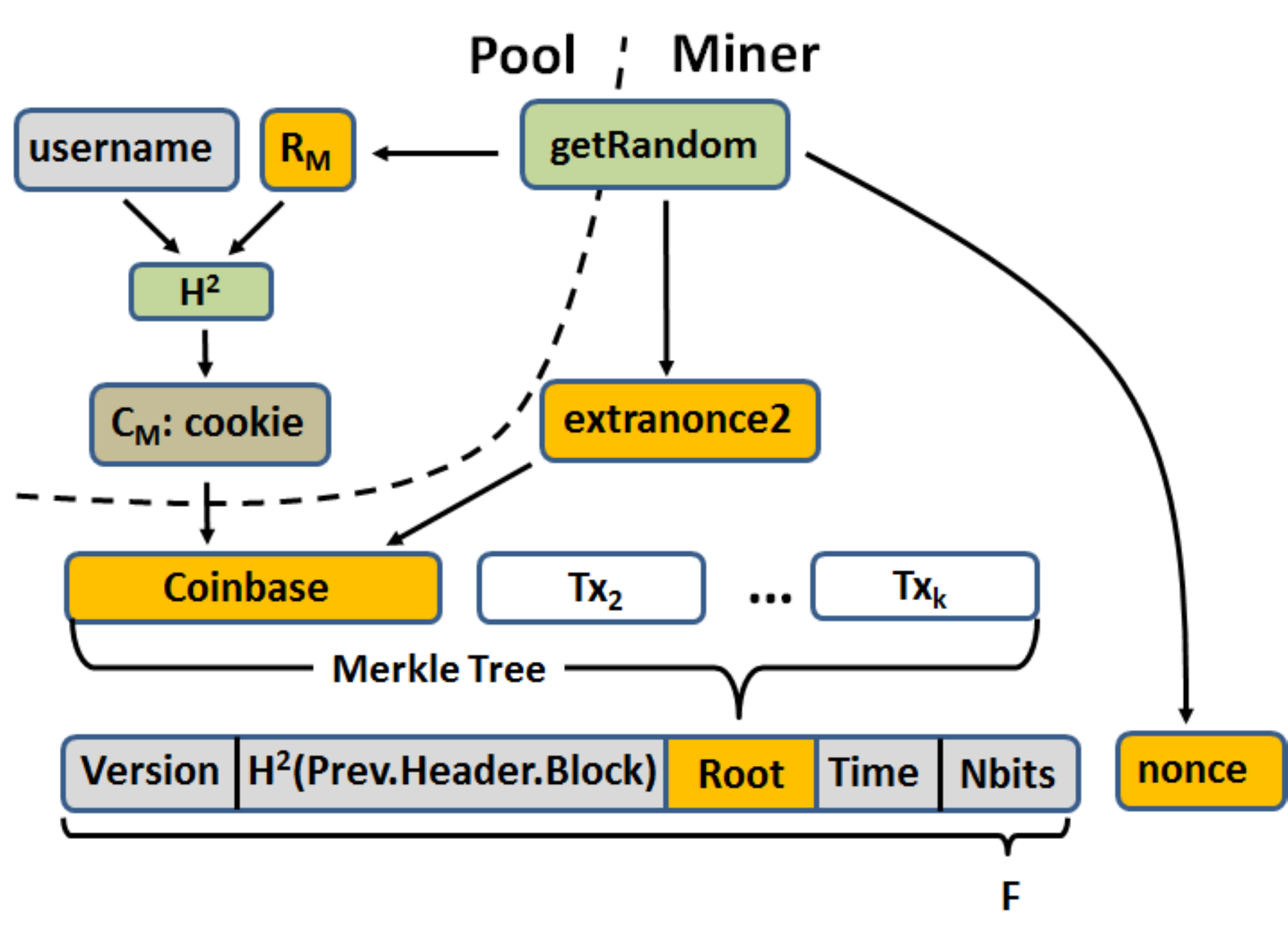}
\vspace{-15pt}
\caption{Bedrock puzzle illustration. The cookie $C_M$ is generated on the
pool, while the $nonce$ and $extranonce2$ are generated on the miner.  The
coinbase transaction contains both $C_M$ and $extranonce2$, see
Figure~\ref{fig:coinbase}.}
\label{fig:bedrock:puzzle}
\vspace{-15pt}
\end{figure}

The share submission packets are particularly vulnerable. First, they can
reveal the target value, thus the difficulty of the jobs on which the miner
works and then the miner's hashrate (see $\S$~\ref{sec:bedrock:discussion}).
Second, share submissions can be hijacked by an active adversary, see
$\S$~\ref{sec:attacks:active}. Encryption of share submissions
will prevent these attacks, but it will strain the pool's resources.

To efficiently address these vulnerabilities, we introduce the concept of {\it
mining cookie}, a secret that each miner shares with the pool, see
Figure~\ref{fig:bedrock:puzzle} and Algorithm~\ref{alg:bedrock}. The miner uses
its mining cookie as an additional, secret field in the Bitcoin puzzle. Without
knowledge of the mining cookie, an adversary cannot infer the progress made by
the miner, thus its hashrate and payout, thus cannot hijack shares submitted by
the miner.

Specifically, let $R_M$ be a random cookie seed that the pool generates for a
miner $M$ Algorithm~\ref{alg:bedrock}, line 3). The pool associates $R_M$ with
$M$, and stores it along with $M$'s symmetric key $K_M$, and its current
$target$ value (line 6). The pool computes $M$'s cookie as $C_M = H^2(R_M,
M.uname)$ (line 4),  where $M.uname$ is the username of the miner. It then
sends $R_M$ to $M$, encrypted with the key $K_M$ (line 7), see
$\S$~\ref{sec:bedrock:solution:set}. The miner similarly uses $R_M$ and its
$username_M$ to compute $C_M$.

To minimally modify Bitcoin, Bedrock stores the cookie as part of the coinbase
transaction (see Figure~\ref{fig:coinbase}), in the place of its unused {\it
previous hash} field. This field is unused since the coinbase transaction does
not have a need for a meaningful input address hash (see
$\S$~\ref{sec:model:coinbase}).  Thus, the puzzle remains the same: The miner
iterates over the $nonce$ and $extranonce2$ values, and reports the pairs that
solve the puzzle, along with its username, in share submission packets (lines
23-26).

To verify the shares, the pool retrieves the miner's key $K_M$, random seed
$R_M$ and $target$ values (line 9).  It uses $R_M$ to reconstruct the cookie
(line 10) and uses $target$, and the reported $nonce$ and $extranonce2$ values,
to reconstruct and verify the puzzle lines 11 ans 12).


\noindent
{\bf Random iterators}.
In the Bitcoin protocol and the Stratum implementation on F2Pool, the $nonce$
and $extranonce2$ values are incremented sequentially: once the miner exhausts
$nonce$, it increments $extranonce2$, then continues to iterate over a reset
$nonce$ value.  In $\S$~\ref{sec:bedrock:discussion} we show that this further
exposes the miner to hashrate inference attacks. We address this problem by
requiring the miner to choose random values for $nonce$ and $extranonce2$ at
each puzzle iteration. To prevent the miner from recomputing an expensive
Merkle tree root at each iteration, we iterate through the $nonce$ space
using a pseudo random permutation (lines 20, 24).

\noindent
{\bf Cookie refresh}.
When a miner mines the current block, i.e., when $H^2(nonce || F || C_M)$ is
less than the target corresponding to the $Nbits$ value, see
$\S$~\ref{sec:model:puzzle}, the puzzle solution needs to be published in the
blockchain.  The published block needs to include all the fields that defined
the puzzle (see $\S$~\ref{sec:model:puzzle}), including the miner's cookie, to
be publicly verified.

To prevent an adversary who monitors the blockchain to learn the
mining cookie of a victim miner and then launch a successful BiteCoin attack
(see $\S$~\ref{sec:bedrock:discussion}), Bedrock changes the mining cookie of
the miner once the miner mines the current block. This is an infrequent
event: for an AntMiner S7 mining equipment, with a hashrate of 4.73 TH/s, and
the current Bitcoin network difficulty (2.58522748405e+11),
Equation~\ref{eq:expected_time_to_share} shows that the expected time to mine a
block is 7.44 years. This is a very low lower bound since it assumes a
constant difficulty. In reality, the difficulty has increased exponentially
since the creation of Bitcoin.
To change the cookie, the pool invokes generateCookie (line 2).


\vspace{-5pt}

\subsubsection{Protect Communicated Secrets}
\label{sec:bedrock:solution:set}

\vspace{-5pt}

Stratum's share difficulty notification messages reveal the difficulty assigned
by the pool to the miner and that the miner uses in the subsequent jobs.
Knowledge of the puzzle difficulty value coupled with the (regulated) share
submission rate, will enable the adversary to infer the hashrate of the miner
(see Equation~\ref{eq:e}), thus its payout.  In addition, Bedrock also needs to
communicate secret values (e.g., the random $R_M$, see
$\S$~\ref{sec:bedrock:solution:cookies}). Bedrock addresses these problems by
extending Stratum's set difficulty notifications to the following {\bf mining
encrypted} message:

\vspace{-10pt}


\[
{\tt mining.encrypted,\ E_{K_M}(param\_list)}
\]

\noindent
where (\textit{param\_list}) is a list of values that need protection, i.e.,
difficulty values and the secret $R_M$. Specifically, \textit{param\_list} can
contain any number of sensitive values in the format {\tt
[[``difficulty'',1024],[``secret'',$R_M$]]}. 


\vspace{-5pt}

\subsubsection{Secure Hashrate Computation}

\vspace{-5pt}

The hashrate inference protocol following a miner subscription and
authorization, as documented in $\S$~\ref{sec:attacks:passive:isp} and
$\S$~\ref{sec:eval:passive:isp} can be exploited also by an adversary to infer
the miner's hashrate. To address this vulnerability, Bedrock requires the miner
to directly report its hashrate during the initial subscription message, along
with other miner capabilities. The miner can locally estimate its hashrate,
e.g., by creating and executing random jobs with a difficulty of 1024. The
miner also needs to factor in its communication latency to the pool, which it
can infer during the subscription protocol.  The miner sends its hashrate
encrypted, using the ``mining encrypted'' message defined above.


If subsequently, the pool receives share submissions from the miner, outside
the desired rate range, it can then adjust the difficulty (through the above
encrypted share difficulty notifications) in order to reflect its more accurate
inference of the miner's hashrate.

\vspace{-5pt}

\section{Discussion}
\label{sec:discussion}

\subsection{Security Discussion}
\label{sec:bedrock:discussion}

\vspace{-5pt}

We now discuss attacks against Stratum and Bedrock, and detail the defenses
provided by Bedrock.

\noindent
\textbf{Target reconstruction attack}.
An attacker that can inspect cleartext subscription response, job assignment
and share submission packets, can reconstruct the job (i.e., puzzle) solved by
the victim miner: Recover $extranonce1$ from an early miner subscription
message, $coinbase1$, $coinbase2$ and the Merkle tree branches from a job
assignment, and $nonce$ and $extranonce2$ from a subsequent share submission
packet. The attacker then reconstructs the $F$ field of the puzzle (see
$\S$~\ref{sec:model:puzzle}) and uses it to infer the miner's hashrate, even
without knowing the puzzle's associated $target$ value. Specifically, the
attacker computes the double hash of $F$ concatenated with $nonce$, then
counts the number of leading zeroes to obtain an upper bound on the job's
target. The attacker then uses recorded inter-share submission time stats and
Equation~\ref{eq:e} to estimate the miner hashrate.

Bedrock thwarts this attack through its use of the cookie $C_M$, a secret known
only by the miner and the pool. The cookie is part of the puzzle. Without its
knowledge, the attacker cannot reconstruct the entire puzzle, thus infer the
target.

\noindent
\textbf{Brute force the cookie}.
The attacker can try to brute force the cookie value. To gain confidence, the
attacker uses the fields from multiple jobs assigned to the same miner to try
each candidate cookie value. A candidate is considered ``successful'' if it
produces a high target value for all the considered jobs. However, in
$\S$~\ref{sec:implementation} we leverage the unused, 256-bit long ``previous
hash'' field of the coinbase transaction, to store the mining cookie. Brute
forcing this field is consider unfeasible.

\noindent
{\bf Resilience to cryptographic failure}.
We assume now an adversary that is able to break the encryption employed by the
pool and the miner, e.g., due to the use of weak random values.  Giechaskiel et
al.~\cite{GCR16} studied the effect of broken cryptographic primitives on
Bitcoin, see $\S$~\ref{sec:related}. While such an adversary can compromise the
privacy of the miner, by recovering the miner's cookie, he will be prevented
from launching active attacks. This is because the miner's cookie is a function
of both a random number and the miner's username.

Specifically, if the attacker hijacks a miner's share submission, the pool
would use the attacker's username instead of the victim's username to construct
the cookie, the coinbase transaction and eventually the header block. The share
will only validate if the attacker managed to find a username that produced a
double hash that was still smaller than the target corresponding to the
difficulty set by the pool.  However, the attacker will need to find such
usernames for each hijacked share.  If the attacker was able to quickly find
such partial collisions, it would be much easier to simply compute the shares
without doing any interception and hijacking.

We further consider an attacker able to break the hash function (invert and
find collisions). Such an attacker can recover a miner's $R_M$ value, then find
a username that produces a collision with the miner's cookie $C_M$.  We observe
however that such an attacker could then be able to also mine blocks quickly,
e.g., by inverting hash values that are smaller than the target corresponding
to the $Nbits$ value.

\vspace{-5pt}

\subsection{Limitations}

\vspace{-5pt}

\noindent
\textbf{Opportunistic cookie discovery}.
When the miner mines the current block, i.e., the double hash of the puzzle is
smaller than the target corresponding to $Nbits$, the miner's cookie is
published in the blockchain.  An adversary who has captured job assignments and
share submissions from the miner, just before this takes place, can use them,
along with the published cookie, to reconstruct the entire puzzle and infer the
miner's hashrate.

This opportunistic attack may take years (e.g., 7.44 years
for an AntMiner S7, see $\S$~\ref{sec:bedrock:solution:cookies}), while, from
our experience, mining equipment has a useful lifetime of around 2 years.
\newmaterial{However, this attack may be more effective against an entity
that owns many homogeneous miners: an adversary may only need days to infer
the rate of a single miner.}

However, to address this limitation, each miner could, at random intervals,
change its operation frequency to a randomly chosen value within an
``acceptable'' operation range. Assuming that the adversary only captures a
limited window of the victim miner's communications, he will only be able to
(i) recover temporary, past hashrate values of the miner, and (ii) reconstruct
the miner's payouts over the monitored interval. Since the miner changes its
operation frequency, once a new cookie is assigned, the adversary will not be
able to predict the miner's future hashrates and payouts.

\noindent
{\bf Verification scope}.
We have only investigated the implementation of Stratum in the pool F2Pool.
However, the identified privacy issues also likely affect other pools, as any
obfuscation to the set difficulty messages would break the compatibility with
the Stratum protocol implemented in current mining equipment.

\vspace{-5pt}

\section{Implementation and Testbed}
\label{sec:implementation}

\vspace{-5pt}

In our experiments, we have used AntMiner S7, a specialized FPGA device for
Bitcoin mining that achieves a hashrate of 4.73 TH/s at 700MHz~\cite{s7_miner}.
We have configured the device for mining on the F2Pool pool, using the Stratum
protocol~\cite{f2pool_help}.

\subsection{Passive Attacks}
\label{sec:implementation:passive}

In order to collect experimental traffic for the passive attacks, we have
leveraged the ability of the AntMiner S7 device to operate at different chip
clock frequencies in order to simulate miner devices with different
capabilities. Specifically, we carried out 24 hour long mining experiments with
the AntMiner S7 operating at frequencies ranging from 100 MHz to 700MHz, with
50MHz granularity. We have used Tcpdump~\cite{jacobson2003tcpdump} to capture
138MB of Stratum traffic of AntMiner S7 devices in the US (May 27 - June
8, 2016) and Venezuela (March 8 - April 2, 2016).  We have sliced the resulting
pcap files into 24 hour intervals using editcap, then processed the
results using python scripts with the scapy library~\cite{biondi2011scapy}.

In addition to the mining traffic, for each of the 24 hour runs, we collected
the empirical payout as reported by the pool, as well as the device hashrate
reported by its internal functionality. We used 24 hour runs because the pool
uses 24 hour cycles for executing payouts. We have manually synchronized the
runs and payout cycles so as to easily correlate the data collected with its
corresponding payout.

\noindent
{\bf StraTap attack}.
To implement the StraTap attack, we have created a script that selects packets
from the captured traces with the ``set\_difficulty'' pattern (invoked method
of the share difficulty notification messages). This pattern signals our script
to perform a share submission count reset, as well as a new recording of the
new difficulty.

\noindent
{\bf ISP Log attack}.
For the ISP Log attack, we used packets sent after the 3-way handshake
initiated by the pool. In addition, to compute more accurate inter-packet
times, we only considered packets that had the PUSH flag set (captured by most
firewall logs, e.g., Snort IDS), thus with non-empty payloads (i.e., no ack
packets that originated on the miner).  The PUSH flag is used to mitigate the
effects of delays on the processing of share submissions, that may end up
causing share rejections. By setting the PUSH flag, miners try to increase the
chance that their shares are quickly processed.

\vspace{-5pt}

\subsection{BiteCoin Attack Implementation}
\label{sec:implementation:bitecoin}

\vspace{-5pt}

\begin{figure}
\centering
\includegraphics[width=0.49\textwidth]{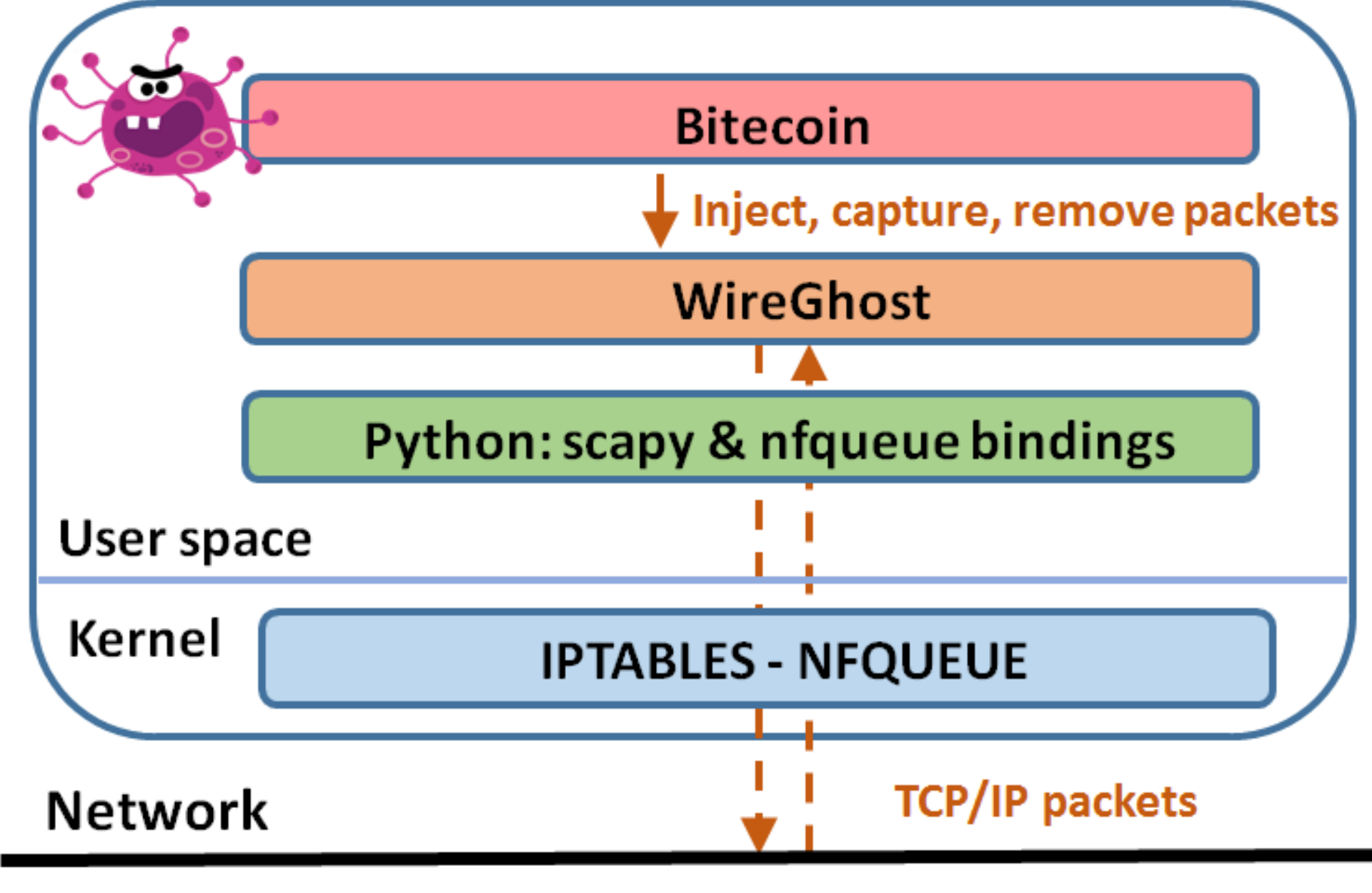}
\caption{Architecture of BiteCoin attack implementation.}
\label{fig:bitecoin:system}
\vspace{-15pt}
\end{figure}

The BiteCoin attack system is illustrated in Figure~\ref{fig:bitecoin:system}.
We have built WireGhost using the iptables nfqueue target, in order to pass
packets into user space. Once it receives network segments in the user space,
it uses the scapy python library to parse and modify packets. Additionally, it
uses the the python nfqueue bindings in order to pass a verdict to the packets.

In order to test BiteCoin and WireGhost, we set up the victim miner behind an
attacker controlled server that performed ``source NAT'' and packet forwarding
for it. This architecture allowed us to emulate an active attacker
intercepting the communication between the miner and the pool. We have
implemented the attacker as a python script that connects to the F2Pool using
Stratum, then intercepts and modifies job assignments and share submissions on
the victim's connection to the pool. While the attacker script does not perform
any mining, in $\S$~\ref{sec:eval:bitecoin} we show that it is able to steal
the victim's hashing power.

\vspace{-5pt}

\subsection{Bedrock Implementation}

\vspace{-5pt}

One requirement of Bedrock is to minimally disrupt the Stratum protocol, see
$\S$~\ref{sec:bedrock:requirements}. Thus, instead of designing the cookie to
be an external field, we seek to leverage unused fields of the coinbase
transaction.  An obvious candidate for the cookie placement is the input script
where the $extranonce1$ and $extranonce2$ reside. However, most pools have
already started using this space for their own internal procedures, e.g., in
F2Pool, to store the miner's name.

Instead, Bedrock uses the yet unused, 32 byte (256 bit) long ``previous input
address'' field of the coinbase transaction, see Figure~\ref{fig:coinbase}.
Since the coinbase transaction rewards the pool with the value of the mined
block (if that event occurs), its input is not used. We have investigated the
Stratum implementation of several pools, including F2Pool~\cite{f2pool_help},
GHash.io~\cite{GHash.io}, SlushPool~\cite{SlushPool} and have confirmed that
none of them use this field. In addition, we note that the size of this field
makes it ideal to store the output of a double SHA-256 hash.

\begin{table}
\centering
\textsf{
\small
\begin{tabular}{l r r}
\textbf{Freq(MHz)} &  \textbf{Hashrate(GHz)}  & \textbf{StraTap Hashrate(GHz)}\\
\midrule
700  &   4720.55 & 4571.48\\
650   &  4371.85 & 4309.96\\
600   &  4040.49 & 4151.27\\
550   &  3693.90 & 3624.13\\
500    & 3365.38 & 3524.57\\
450   &  3030.01 & 3154.80\\
400  &   2689.34 & 2696.72\\
350    & 2364.61 & 2382.17\\
300    & 2023.65 & 2039.55\\
250    & 1687.17 & 1699.91\\
200    & 1347.14 & 1274.29\\
150    & 1010.19 & 1007.06\\
100   &  672.55 & 703.28\\
\end{tabular}
}
\caption{Operation frequency, actual hashrate and StraTap inferred hashrate.
We observe the correlation between the actual and the average hashrate, that
allowed StraTap to achieve a good payout estimate.
%
}
\label{table:freq_hashrate_payout}
\vspace{-15pt}
\end{table}


\vspace{-5pt}

\section{Evaluation}
\label{sec:eval}

\vspace{-5pt}

In this section we evaluate the StraTap, ISP Log and BiteCoin attacks, as well
as the performance of Bedrock. We use the mean squared error (MSE) and the mean
percentage error (MPE) to evaluate the accuracy of the predictions made by the
passive attacks. Specifically, let $P = \{ P_1, .., P_n\}$ be a set of observed
daily payments over $n$ days, and let $\bar{P} = \{ \bar{P_1}, .., \bar{P_n}
\}$ be the corresponding predicted daily payments for the same days. Then,
MSE($\bar{P},P$) = $\frac{1}{n} \sum_i^n (\bar{P_i} - P_i)^2$, and
MPE($\bar{P},P$) = $\frac{100\%}{n} \sum_i^n \frac{P_i - \bar{P_i}}{P_i}$.

\vspace{-5pt}

\subsection{The StraTap Attack}
\label{sec:eval:passive:stratap}

\vspace{-5pt}

\begin{figure}[t]
\centering
\includegraphics[width=0.45\textwidth]{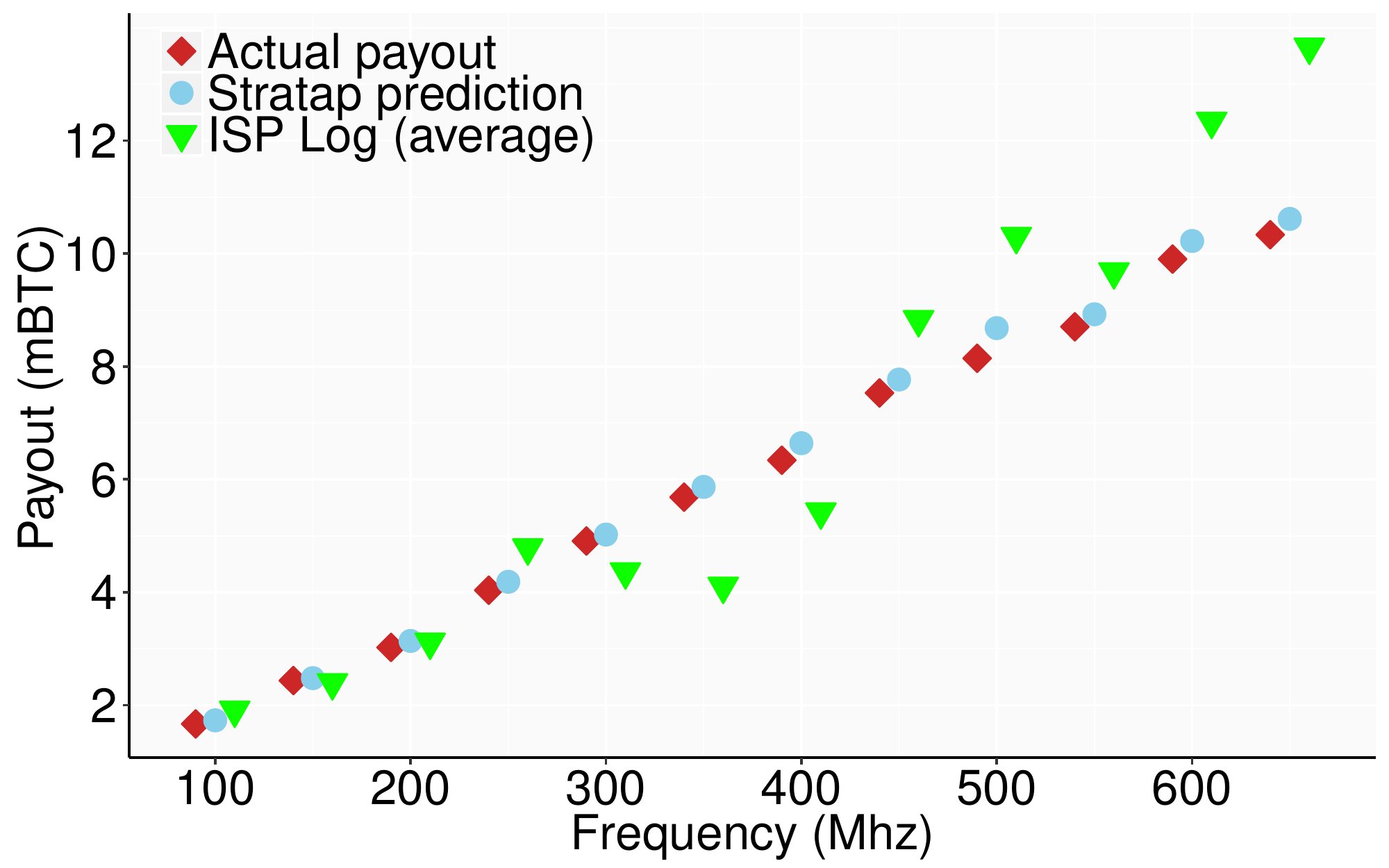}
\caption{Payout prediction by StraTap and ISP Log attacks, compared to
empirical payout, in mili Bitcoin (mBTC), as a function of the miner's
frequency of operation (MHz). The \textit{actual payout} series (red diamonds)
corresponds to daily payouts collected from the F2Pool account records. The
StraTap payout series (blue disks) shows daily payout predictions based on
entire Stratum messages intercepted. The ISP Log series (green triangles) shows
the daily payout prediction when using the average inter-packet times over 50
packets. StraTap's prediction error ranges between 1.75-6.5\% (MSE=0.062,
MPE=3.46\%). ISP Log has an error between 0.53 - 34.4\% (MSE = 2.02, MPE =
-9.49\%).}
\label{fig:eval:stratap}
\vspace{-15pt}
\end{figure}

We have used the StraTap script described in
$\S$~\ref{sec:implementation:passive} to calculate the average time of share
creation for each of the detected intervals of constant difficulty. For each of
the 24 hour runs, we also calculated the weighted average difficulty as well as
the weighted average hashrate for the entire run. In addition, we have also
used Equation~\ref{eq:e}, along with the computed average time and recorded
difficulty values, to compute a prediction of the weighted average hashrate of
the miner.

Table~\ref{table:freq_hashrate_payout} shows the AntMiner's frequency of
operation, the output hashrate achieved at that frequency, and the predicted
hashrate. As expected, there is a linear relationship between the frequency of
operation and the device's hashrate achieved. As a consequence, this
relationship is preserved across the empirical payout reported by the pool
operators.

Specifically, we have used the pool's hashrate to BTC conversion (see
$\S$~\ref{sec:stratum}) to predict the miner's resulting daily payout.
Figure~\ref{fig:eval:stratap} shows the data series for the empirical and
predicted payouts, versus the operation frequency of the miner.  The StraTap
attack achieves a prediction error of between 1.75\% and 6.5\%, with a mean
square error (MSE) of 0.062 and mean percentage error (MPE) of -3.46\%. Thus,
StraTap's predictions tend to be slightly larger than the actual payout values.

\subsection{The ISP Log Attack}
\label{sec:eval:passive:isp}

\vspace{-5pt}

\begin{figure}[t]
\centering
\subfigure[]
{\label{fig:timeline:hashrate:200MHz}{\includegraphics[width=0.49\textwidth]{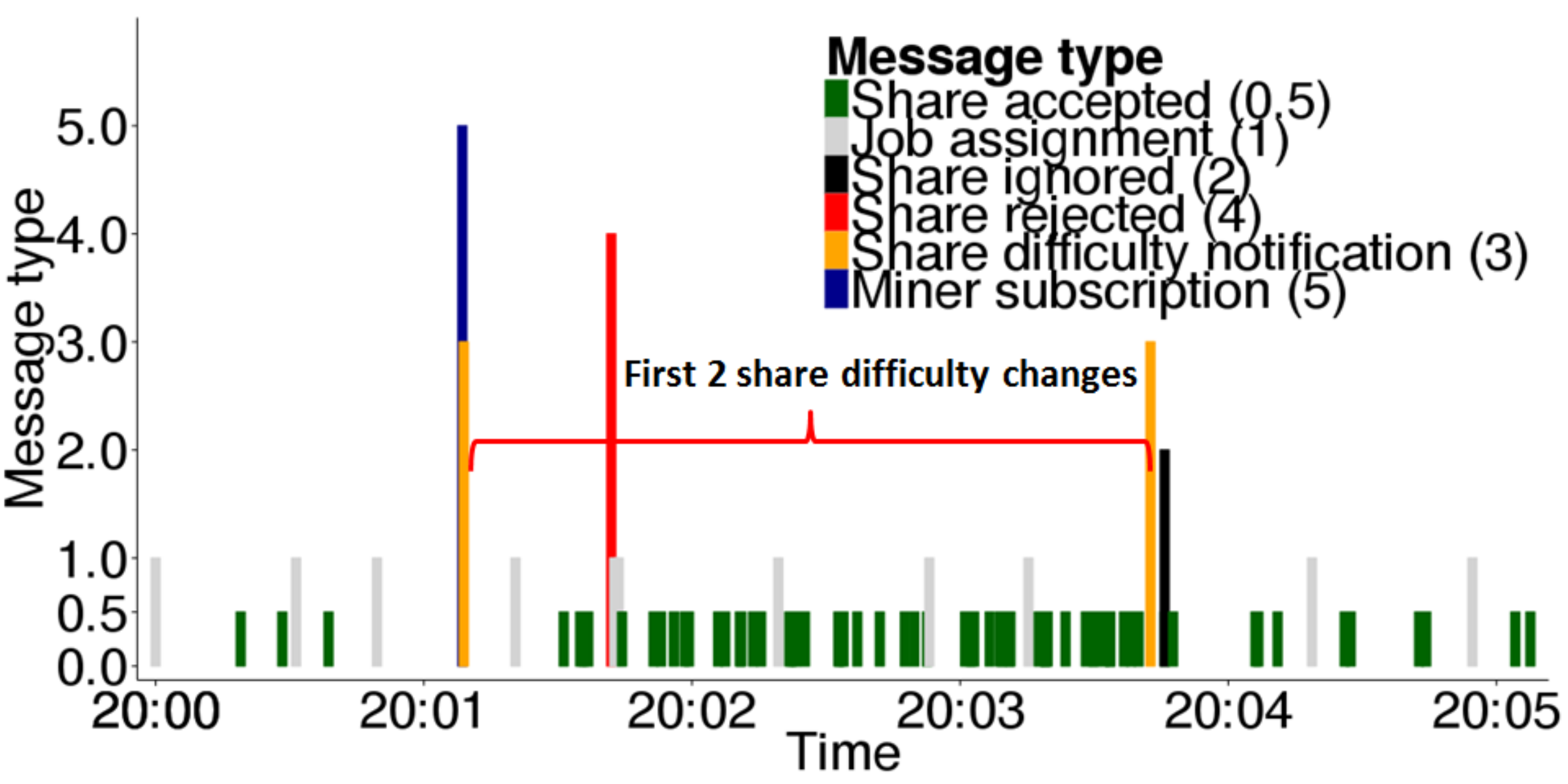}}}
\vspace{-5pt}
\subfigure[]
{\label{fig:timeline:hashrate:600MHz}{\includegraphics[width=0.49\textwidth]{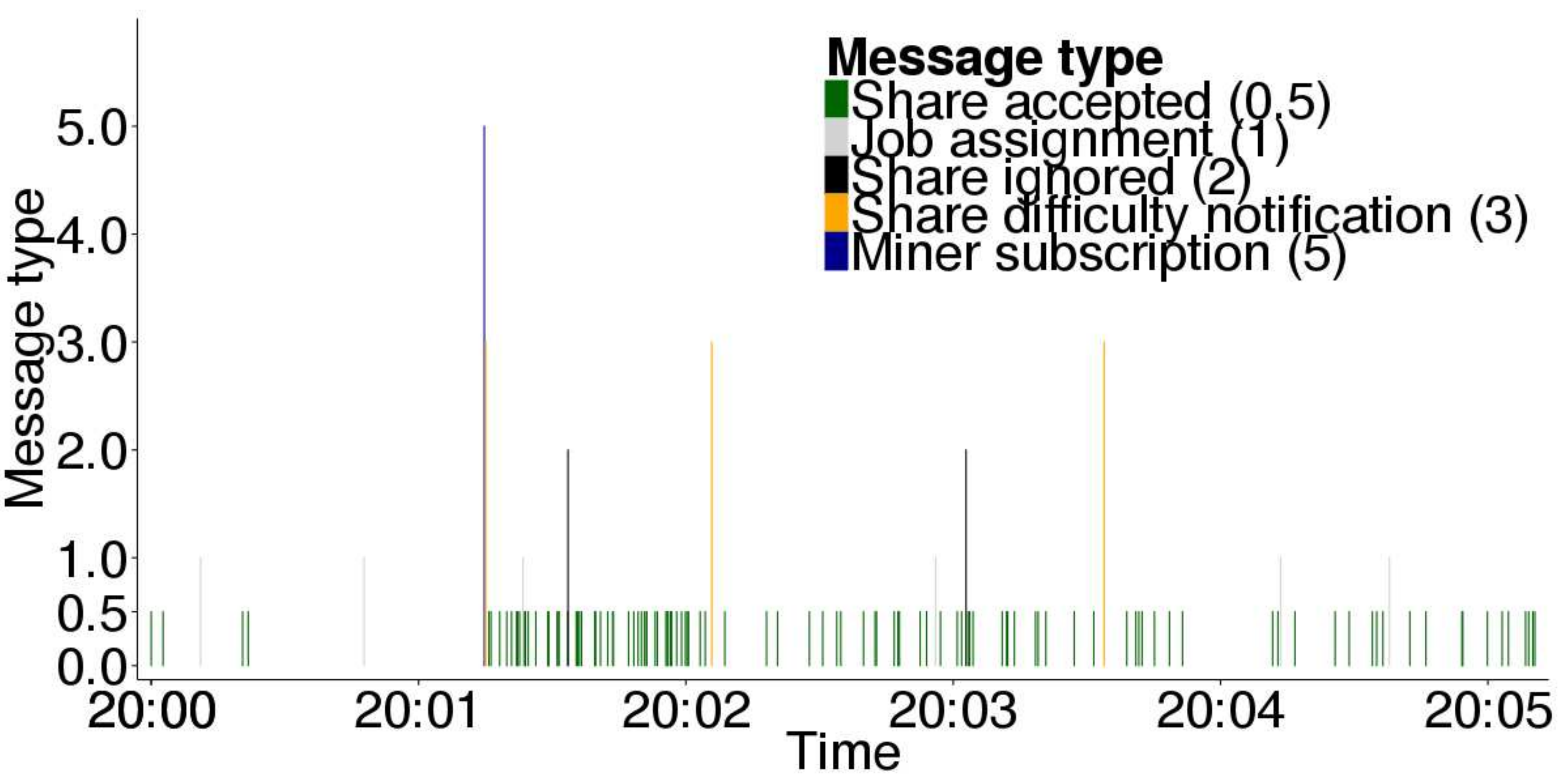}}}
\vspace{-5pt}
\caption{Timelines that focus on the interval between the first two share
difficulty notifications, following a miner subscription and authorization
protocol, when
(a) the miner operates at 200MHz and
(b) the miner operates at 600MHz.
While the intervals between the first two such notifications at both
frequencies contain approximately 50 share submission packets, this interval is
significantly shorter at 600MHz. This is because at 600MHz the miner can solve
the 1024 difficulty puzzles much faster than at 200MHz.  The ``ISP Log'' attack
exploits this observation to infer the hashrate of the miner, while only
counting packets (i.e., without being able to inspect them).}
\label{fig:timeline:hashrate}
\vspace{-15pt}
\end{figure}

We first present results of our analysis of F2Pool's hashrate inference
protocol. We then show the ability of the ISP Log attack to leverage these
findings to infer the miner's daily payouts, given only metadata of the miner's
packets.

\begin{table}
\centering
\textsf{
\small
\begin{tabular}{l r r}
\textbf{Freq(MHz)} &  \textbf{\# of Packets}  & \textbf{Time Interval}\\
\midrule
100 & 57 & 288.872897148\\
150 & 56 & 256.145660877\\
200 & 51 & 153.622557878\\
250 & 63 & 146.007184982\\
300 & 55 & 131.089562893\\
350 & 62 & 146.259056807\\
400 & 54 & 101.954112053\\
450 & 67 & 104.665092945\\
500 & 50 & 58.2229411602\\
550 & 62 & 76.0586118698\\
600 & 54 & 50.7432210445\\
650 & 56 & 45.6691811085\\
\end{tabular}
}
\caption{Number of share submission packets for the initial $1024$ difficulty
period, as well as the length of the time interval when the pool accepted those
shares, for various miner frequencies of operation. At any miner operation
frequency, at least $50$ share submission packets are accepted, irrespective of
wait time.  This process enables the pool and the ISP Log attack to infer the
miner's hashrate.
}
\label{table:timeline_1024}
\vspace{-15pt}
\end{table}

\noindent
{\bf Hashrate inference protocol}.
As mentioned in $\S$~\ref{sec:attacks:passive:isp}, immediately following the
miner subscription and authorization, the pool sets the difficulty to 1024, and
changes it only after receiving a sufficient number of share submissions to
infer the miner's hashrate. For instance,
Figure~\ref{fig:timeline:hashrate:200MHz} shows that when the miner operates at
200MHz, the number of share submissions between the first two share difficulty
notification messages is similar to the number of share submissions when the
miner operates at 600MHz (Figure~\ref{fig:timeline:hashrate:600MHz})
(approximately 50).  However, the time interval between the first two share
difficulty notifications is much shorter at 600MHz: the miner can compute 50
shares at the constant difficulty 1024 much faster than when operating at
200MHz.

More general, Table~\ref{table:timeline_1024} shows the number of share
submission packets sent for this initial measurements period for each of the
frequencies analyzed.  We observe that the pool requires that this process
generates at least $50$ share submissions, irrespective of the miner operation
frequency. The pool waits up to 288 seconds to receive the required number of
shares, before sending the second share difficulty notification.

We conjecture that the pool uses this process in order to infer the hashrate of
the miner, which it needs in order to assign jobs (puzzles) that a miner can
solve at a ``desirable'' rate. Specifically, large pools handle thousands of miners
simultaneously~\footnote{The Bitcoin network currently has around 100,000
miners~\cite{MinerCount,Corti}, of which at least 16\% work with
F2Pool~\cite{F2PoolShare}.}. In order to minimize the time it takes to process
share submissions received from thousands of miners, the pool
needs to regulate the rate at which a miner submits shares, irrespective of the
miner's computing capabilities. Figure~\ref{fig:timeline:rate} illustrates
this share submission rate control. In our experiments we observed that for F2Pool,
this rate ranges to between 1 to 4 share submissions per minute. A second reason
for this process stems from the need of miners to prove computation progress
and gain regular payouts.

\begin{figure}[t!]
\vspace{-15pt}
\centering
\includegraphics[width=0.43\textwidth]{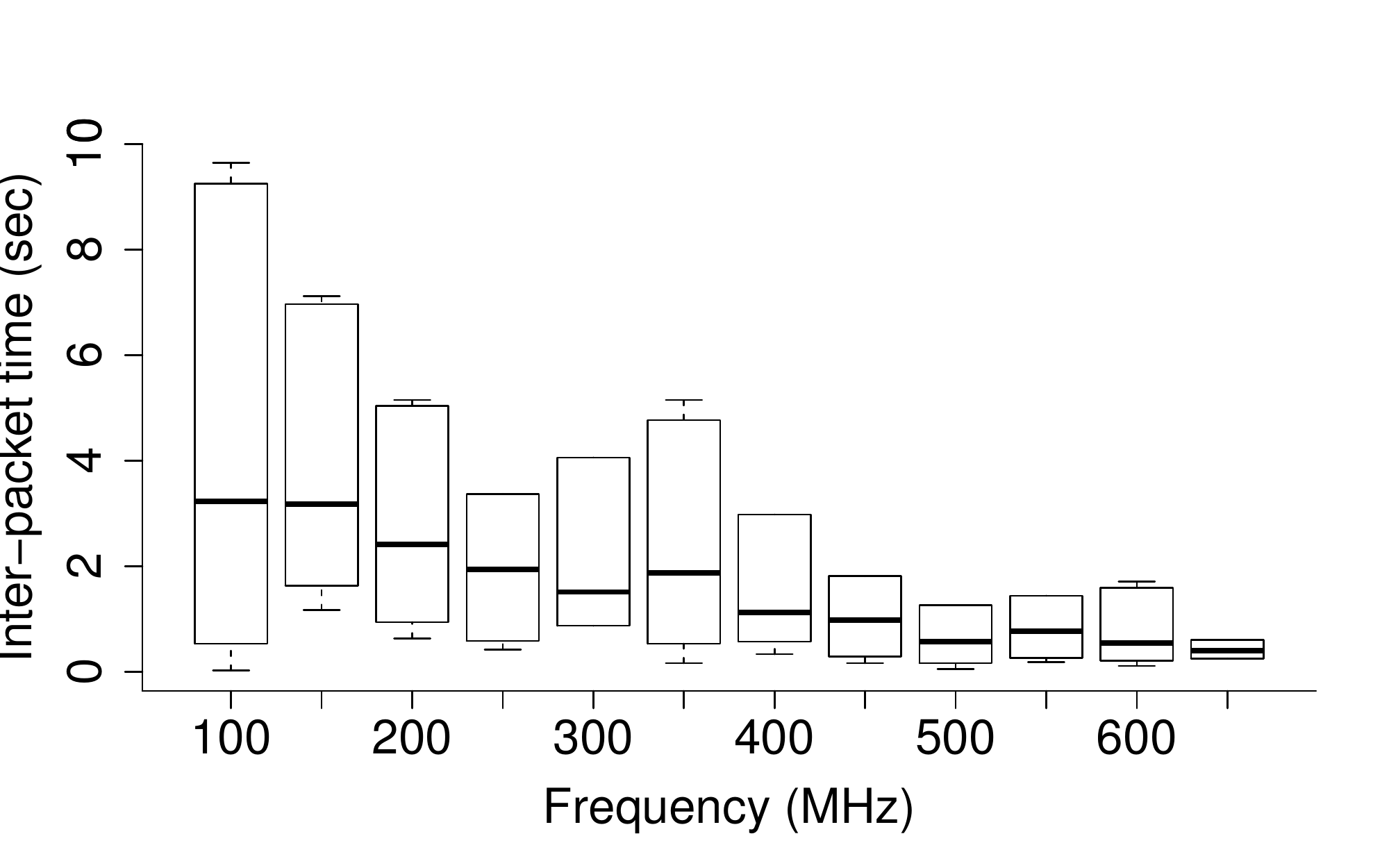}
\vspace{-5pt}
\caption{
1st, 2nd and 3rd quartile for the inter-packet times of the first 50
packets during the initial difficulty setting procedure, as a function of the
miner's operating frequency. We observe a monotonically decreasing tendency
of the inter-packet times, with an increase in the miner capabilities.
This suggests that inter-packet time stats over the first 50 packets can provide
a good hashrate estimator for the ISP Log attack.
\label{fig:isp:log:predictor}}
\vspace{-15pt}
\end{figure}

\noindent
{\bf ISP Log attack results}.
We have implemented the ISP Log attack using statistics of the inter-packet
arrival time of the first 50 packets sent by the miner to the pool, after a
detected 3-way miner subscription and authorization protocol.
%
%
Figure~\ref{fig:isp:log:predictor} shows the 1st, 2nd (median) and 3rd
quartiles of the inter-packet times, for the first 50 packets, when the miner
operates at frequencies ranging from 100 to 650 MHz. The linearly decreasing
behavior of the median, 1st and 3rd quartiles indicates that statistics over
the inter-packet times of the first 50 packets, may make a good predictor.



To confirm this, we have used the mean inter-packet time over the first 50
packets to predict the miner's hashrate and then its payout.
Figure~\ref{fig:eval:stratap} compares the ISP Log attack daily payout
prediction with that of StraTap and with the empirical payout. The ISP Log has
an error that ranges between 0.53\% and 34.4\%, with an MSE of 2.02 and MPE of
-9.49\% . Thus, ISP Log over predicts the daily payouts, and, as expected, it
exceeds the error of the StraTap attack. 

\vspace{-5pt}

\subsection{BiteCoin: Proof of Concept}
\label{sec:eval:bitecoin}

\vspace{-5pt}

\begin{figure}[t]
\centering
\subfigure[]
{\label{fig:bitecoin:timeline:attacker}{\includegraphics[width=0.49\textwidth]{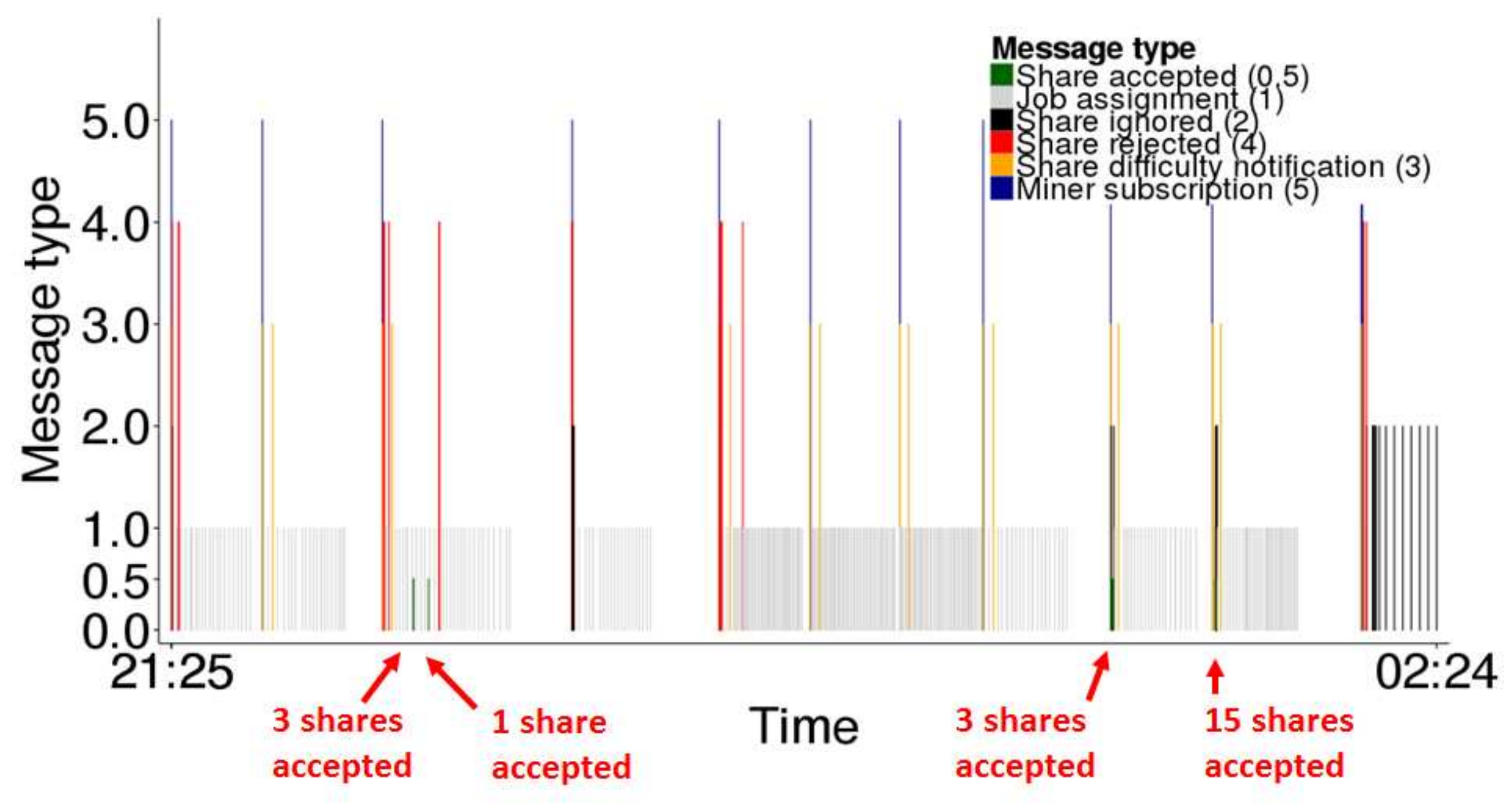}}}
\vspace{-5pt}
\subfigure[]
{\label{fig:bitecoin:timeline:victim}{\includegraphics[width=0.49\textwidth]{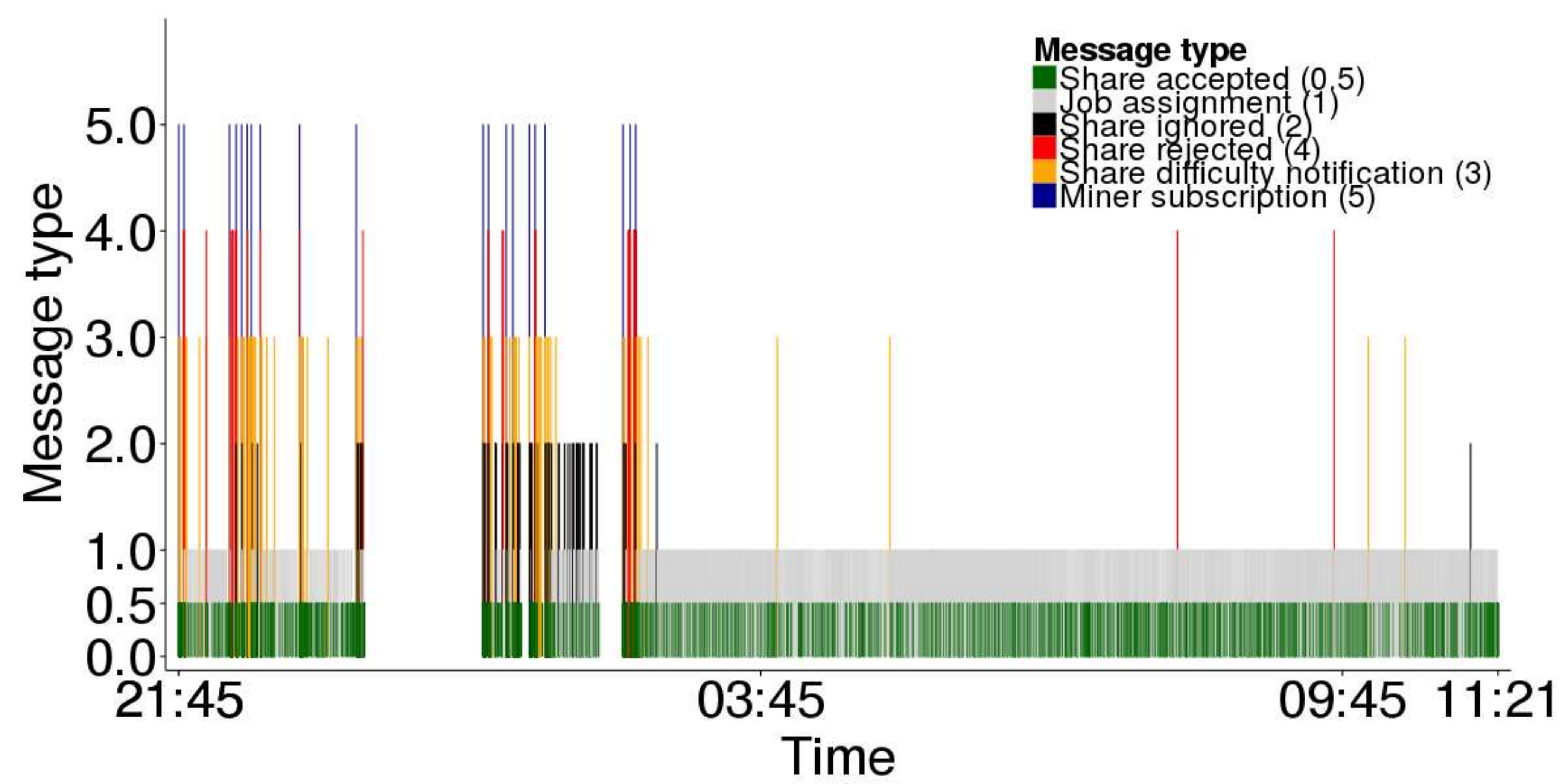}}}
\vspace{-5pt}
\caption{Greedy BiteCoin attack timelines for
(a) adversary and (b) victim miner. In a 5h interval, the attacker hijacked 342
job assignments and 72 corresponding share submissions of the victim miner.
23 shares (the green clumps marked with red arrows) were accepted by the pool.}
\label{fig:bitecoin:timeline}
\vspace{-15pt}
\end{figure}

We have experimented with the BiteCoin implementation described in
$\S$~\ref{sec:implementation:bitecoin}. Specifically, the attacker greedily
injected all the jobs assigned by the pool into the victim communication stream
during the attack time and without any modification. Our implementation
injected a total of 342 job assignments in a time interval of 5 hours, from
hour 21:25 to 02:24. The attacker monitored the share submissions from the
victim, and hijacked shares corresponding to the injected jobs.

Figure~\ref{fig:bitecoin:timeline} shows the results of this attack. The
adversary, whose timeline is shown in
Figure~\ref{fig:bitecoin:timeline:attacker}, hijacked 72 share submissions from
the victim miner. 23 shares (the green clumps marked with red arrows) were
accepted by the pool, i.e., as if they were mined by the attacker and not by
the victim. 49 shares were rejected. Figure~\ref{fig:bitecoin:timeline:victim}
shows the timeline of the attack from the perspective of the victim miner.

The gaps are likely due to the script trying to get some constant work in.
Every disconnection and reconnection of the attacker will trigger a subscribe
protocol where the first job has the true flag set. This would explain why
there are no hijacked shares between around 22:00 and 1:00 in the attacker
timeline and also the gap of any activity in the victim timeline. These
constant reconnects may have constantly blanked the job pool of the victim
until the attacker was able to maintain its connection to submit the shares.


\vspace{-5pt}

\subsection{The Bedrock Evaluation}
\label{sec:eval:bedrock}

\vspace{-5pt}

\begin{figure}
\centering
\includegraphics[width=0.45\textwidth]{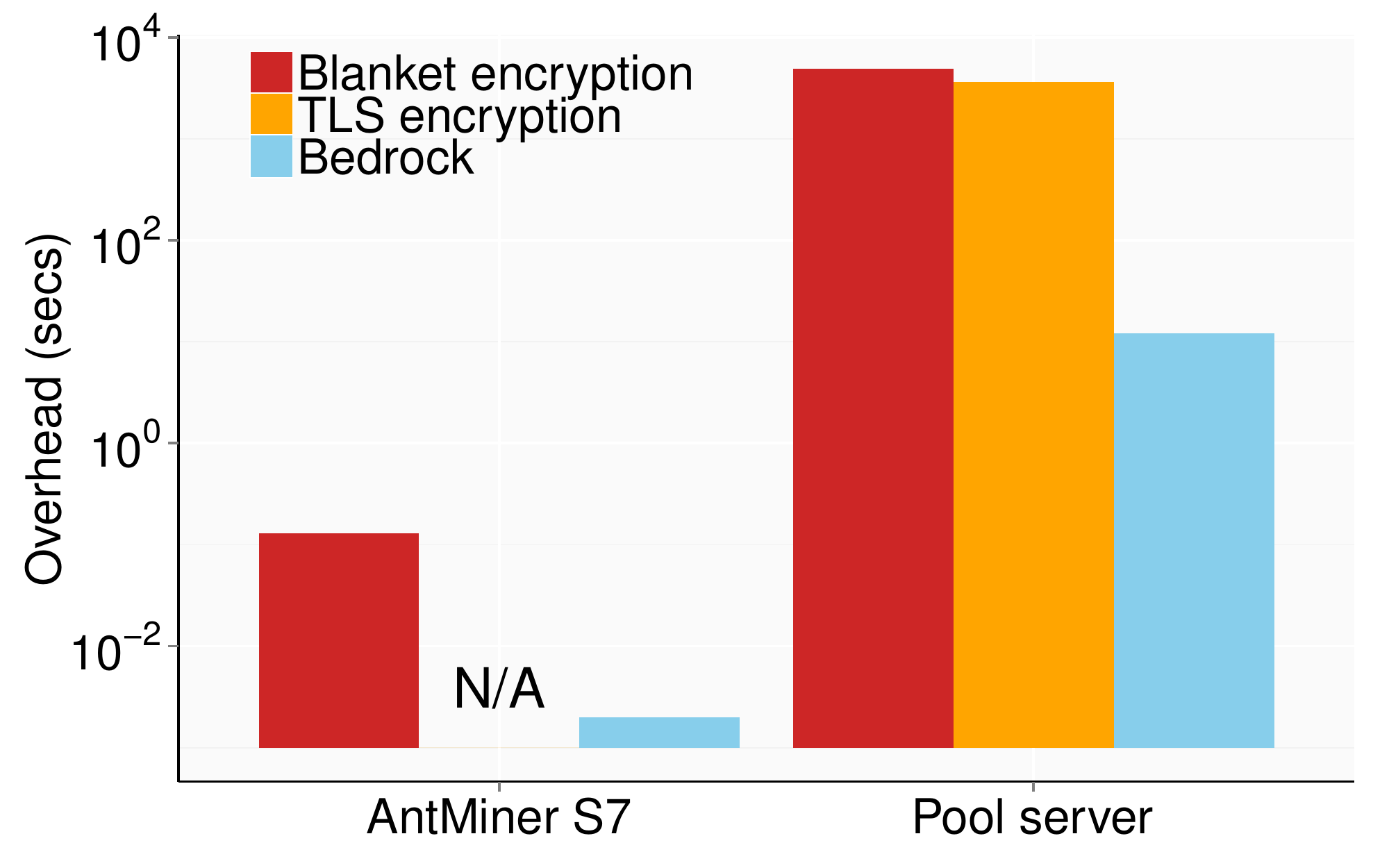}
\caption{Overhead comparison of Bedrock and a complete encryption approach, for
miner and pool. Bedrock imposes a small daily overhead on both the pool (12.03s
to handle 16,000 miners) and miner (0.002s). However, a solution that
encrypts all Stratum packets imposes a daily overhead of 1.36 hours on the
pool.}
\label{fig:eval:bedrock}
\vspace{-15pt}
\end{figure}

We measured Bedrock's encryption times when using AES-256 in CBC mode on the
AntMiner S7 and on a server with 40 cores Intel(R) Xeon(R) CPU E5-2660 v2 @
2.20GHz and 64 GB RAM. The AntMiner was able to encrypt 1024-blocks at
32,231.09 Kb/sec while the server was able to encrypt at 86,131.07 Kb/sec for
the same block size.


Based on the collected data, Stratum generates an average of 31.63 set
difficulty messages per day. Figure~\ref{fig:eval:bedrock} shows that Bedrock
imposes a 0.002s decryption overhead per day on an AntMiner S7, while on a pool
using the above server to handle 16,000 miners, it imposes an encryption
overhead of 12.03 seconds per day.

In contrast, a solution that encrypts each Stratum packet imposes an overhead
of 0.13 seconds per day on the AntMiner, and an unacceptable 1.36 hours per day
on the pool server, to handle 16,000 miners.

\vspace{-15pt}

\subsubsection{TLS Overheads}

\newmaterial{
We also compare Bedrock against Stratum protected with TLS. We have used a
replay of a 24 hour subset of our Stratum traffic dataset
($\S$~\ref{sec:implementation:passive}), sent over TLS between a laptop used as
miner (AntMiner does not support TLS) and the server above, used as the pool.}

\noindent
\newmaterial{
{\bf Computation overheads}.
To measure the TLS computation overheads, we have used
Tcpdump~\cite{jacobson2003tcpdump} to capture the times when Stratum/TLS
packets leave from and arrive at the pool application, and also captured the
time when the packets are sent from/received by the pool TLS socket.  We have
computed the total daily pool side TLS overhead of sending and receiving
Stratum packets (job assignment, share submission, notifications, set
difficulty change, etc). Figure~\ref{fig:eval:bedrock} shows the difference
between this overhead and the same overhead but when using bare TCP. It shows
that the daily computation overhead imposed by TLS on the pool, through the
traffic of 16,000 miners, is 1.01 hours.} This amounts to a
computational overhead percentage of at least 4.3\%.


\noindent
\newmaterial{
{\bf Bandwidth overhead}.
In addition, we have measured the bandwidth overhead imposed by TLS. The total
miner-to-pool payload (single miner) for cleartext Stratum/TCP traffic is
465,875 bytes and for Stratum/TLS is 738,873 bytes. The total pool-to-miner
payload of Stratum/TCP is 3,852,795 bytes while for Stratum/TLS is 4,062,956
bytes. Thus, TLS imposes a 58\% overhead on the miner-to-pool bandwidth, for a
total of 4.05GB daily overhead on the pool from 16,000 miners. This uplink
overhead is significant, especially for miners in countries with poor Internet
connectivity.}

\newmaterial{
TLS imposes a 5\% overhead on the pool-to-miner bandwidth, for a total of
3.13GB daily overhead on the pool. The TLS overhead is much larger in
miner-to-pool communications, even though there are more pool-to-miner packets.
This is because the miner-to-pool share submission packets are much smaller
than the pool-to-miner job assignments, thus the TLS overhead (125 to 160
bytes) becomes a significant factor for them.} In contrast, the
percentage bandwidth overhead for Bedrock is only 0.04\%.

\noindent
\newmaterial{
{\bf Conclusions}.
Bedrock is more efficient than blanket encryption and TLS. While the pool could
use more equipment to handle encryption more efficiently, blanket encryption
and TLS do not address the hashrate inference vulnerability. In addition, TLS
imposes a significant uplink bandwidth overhead on miners.}

\vspace{-10pt}

\section{Related Work}
\label{sec:related}

\vspace{-5pt}

\noindent
{\bf Bitcoin mining attacks}.
Decker and Wattenhofer~\cite{DW13} study Bitcoin's use of a multi-hop broadcast
to propagate transactions and blocks through the network to update the ledger
replicas, then study how the network can delay or prevent block propagation.
Heilman et al.~\cite{HKZG15} propose eclipse attacks on the Bitcoin network,
where an attacker leverages the reference client's policy for updating peers to
monopolize all the connections of a victim node, by forcing it to accept only
fraudulent peers. The victim can then be exploited to attack the mining and
consensus systems of Bitcoin.  Bissas et al.~\cite{BLOAH16} present and
validate a novel mathematical model of the blockchain mining process and use it
to conduct an economic evaluation of double-spend attacks, both with and
without a concurrent eclipse attack.

Courtois and Bahack~\cite{CB14} propose a practical block withholding attack,
in which dishonest miners seek to obtain a higher reward than their relative
contribution to the network. They also provide an excellent background
description of the motivation and functionality of mining pools and the mining
process.

\noindent
{\bf Bitcoin anonymity}.
Significant work has focused on breaking the anonymity of Bitcoin
clients~\cite{BKP14,KKM14,MPJLMVS13,AKRSC13}. For instance, Biryukov et
al.~\cite{BKP14} proposed a method to deanonymize Bitcoin users, which allows
to link user pseudonyms to the IP ad- dresses where the transactions are
generated. Koshy et al.~\cite{KKM14} use statistical metrics for mappings of
Bitcoin to IP addresses, and identify pairs that may represent ownership
relations.

Several solutions arose to address this problem. Miers et al.~\cite{MGGR13}
proposed ZeroCoin, that extends Bitcoin with a cryptographic accumulator and
zero knowledge proofs to provide fully anonymous currency transactions.
Ben-Sasson et al.~\cite{BSCG0MTV14} introduced Zerocash, a decentralized
anonymous payment solution that hides all information linking the source and
destination of transactions. Bonneau et al.~\cite{BNMCKF14} proposed
Mixcoin, a currency mix with accountability assurances and randomized fee
based incentives.

Our work is orthogonal to previous work on Bitcoin anonymity, as it identifies
vulnerabilities in Stratum, the communication protocol employed by
\newmaterial{cryptocurrency} mining solutions. As such, our concern is for the
privacy and security of the miners, as they generate coins. Our attacks are
also more general, as they apply not only to Bitcoin, but to a suite of other
popular altcoin solutions,
e.g.,~\cite{litecoin_stratum,ethereum_stratum,monero_stratum} that build on
Stratum.

\noindent
{\bf Effects of broken crypto on Bitcoin}.
Giechaskiel et al.~\cite{GCR16} systematically analyze the effects of broken
cryptographic primitives on Bitcoin. They reveal a wide range of possible
effects that range from minor privacy violations to a complete breakdown of the
currency. Our attacks do not need broken crypto to succeed.  However, we show
that Bedrock, our secure Stratum extension is resilient to broken crypto
primitives.

\vspace{-10pt}

\section{Conclusions}

\vspace{-5pt}

In this paper we have shown that the lack of security in Stratum, Bitcoin's
mining communication protocol, makes miners vulnerable to a suite of passive
and active attacks, that expose their owners to hacking, coin and equipment
theft, loss of revenue, and prosecution. We have implemented and shown that the
attacks that we introduced are efficient. Our attacks reveal that encryption is
not only undesirable, due to its significant overheads, but also ineffective:
an adversary can predict miner earnings even when given access to only the
timestamps of miner communications. We have developed Bedrock, a minimal and
efficient Stratum extension that protects the privacy and security of mining
protocol participants. We have shown that Bedrock imposes an almost negligible
computation overhead on the mining participants and is resilient to active
attacks even if the used cryptographic tools are compromised.

\section{Acknowledgments}

\newmaterial{
We thank the shepherds and the anonymous reviewers for their excellent
feedback.  We thank Patrick O'Callaghan for suggesting this problem and for
insightful discussions. This research was supported by NSF grants 1526494 and
1527153.}

\vspace{-10pt}

\bibliographystyle{unsrt}
\bibliography{bitcoin}

\end{document}